\documentclass[aps,pre,twocolumn,floatfix,showpacs,superscriptaddress,groupedaddress]{revtex4}
\usepackage{amsmath, amsthm, amssymb}
\usepackage{mathtools}
\usepackage{graphicx, graphics}
\usepackage{epsfig}
\usepackage{times}
\usepackage{verbatim}   
\usepackage{color} 
\usepackage{subfigure}
\usepackage{dcolumn}
\usepackage{float}
\usepackage{enumerate}
\DeclarePairedDelimiter{\ceil}{\lceil}{\rceil}

\newcommand{\fig}{Fig.}

\newcommand{\sect}{Sec.\ }
\newcommand{\eqn}{Eq.\ }
\newcommand{\eqns}{Eqs.\ }

\begin{document}

\title{Pulse-coupled model of excitable elements on heterogeneous sparse networks}

\author{P. Piedrahita}
\email{ppiedrahita@gmail.com}
\affiliation{Departamento de F\'isica de la Materia Condensada, Universidad de Zaragoza, Zaragoza 50009, Spain}
\affiliation{Instituto de Biocomputaci\'on y F\'isica de Sistemas Complejos, Universidad de Zaragoza, Zaragoza 50018, Spain}
\author{J. J. Mazo}
\affiliation{Departamento de F\'isica de la Materia Condensada, Universidad de Zaragoza, Zaragoza 50009, Spain}
\affiliation{Instituto de Biocomputaci\'on y F\'isica de Sistemas Complejos, Universidad de Zaragoza, Zaragoza 50018, Spain}
\author{L. M. Flor\'{\i}a}
\email{mario.floria@gmail.com }
\affiliation{Departamento de F\'isica de la Materia Condensada, Universidad de Zaragoza, Zaragoza 50009, Spain}
\affiliation{Instituto de Biocomputaci\'on y F\'isica de Sistemas Complejos, Universidad de Zaragoza, Zaragoza 50018, Spain}
\author{Y. Moreno}
\affiliation{Instituto de Biocomputaci\'on y F\'isica de Sistemas Complejos, Universidad de Zaragoza, Zaragoza 50018, Spain}
\affiliation{Departamento de F\'isica Te\'orica, Universidad de Zaragoza, Zaragoza 50009, Spain}


\begin{abstract}
	We study a pulse-coupled dynamics of excitable elements in uncorrelated scale-free networks. Regimes of self-sustained activity are found for homogeneous and inhomogeneous couplings, in which the system displays a wide variety of behaviors, including periodic and irregular global spiking signals, as well as coherent oscillations, an unexpected form of synchronization. Our numerical results also show that the properties of the population firing rate depend on the size of the system, particularly its structure and average value over time. However, a few straightforward dynamical and topological strategies can be introduced to enhance or hinder these global behaviors, rendering a scenario where signal control is attainable, which incorporates a basic mechanism to turn off the dynamics permanently. As our main result, here we present a framework to estimate, in the stationary state, the mean firing rate over a long time window and to decompose the global dynamics into average values of the inter-spike-interval of each connectivity group. Our approach provides accurate predictions of these average quantities when the network exhibits high heterogeneity, a remarkable finding that is not restricted exclusively to the scale-free topology.
\end{abstract}


\maketitle

\section{Introduction}

Neural models have a very long history. The first mathematical model of a neuron was proposed by a pioneer named Louis Lapicque in 1907 \cite{lapicque} and, for the most part of the XX century, just a handful of remarkable researchers contributed to this field with abundant discoveries. Among these models and theories, those that attempt to capture intricate mechanisms of real neurons and the interaction among them stand out. To mention a few: The synaptic plasticity introduced by Donald Hebb (1940s), the spike generation mechanism by Alan Hodgkin and Andrew Huxley (1950s), the cable theory by Wilfrid Rall (1950s), the interacting neurons by David Marr (1960s-1970s) and the working memory studies by Daniel Amit (1980s-1990s) are some of the most significant developments in Theoretical Neuroscience during this period \cite{Hebb1949,hodgkin1952,rall1962,rall1964,marr1969,Amit1997a,Amit1997b}.\\
From the decades of 1980s and 1990s, with the massive introduction of computers to scientific research, there has been a continuous growing interest on the study of neural systems at many different levels and scales. This phenomenon has led to the creation of new fields of investigation that are extremely active nowadays. With computers now it is possible to perform extensive simulations of large numbers of firing neurons, we can generate diverse underlying structures of interactions and include many realistic ingredients like chemical synapses, ion currents, compartments, stochastic currents, noise and adaptative parameters \cite{dayan2001,haken2007,gerstner2002,gerstner2014}. As for the interactions among neurons, networks have become a standard framework when modeling neural systems \cite{Boccaletti2006,Barrat2008,Newman2010}. A great deal of research has been devoted specifically to study emergent properties of ensembles of neurons that interact according to both simple and complex topologies \cite{Mirollo1990,corral95prl,Roxin2004,Eguiluz2005,Piedrahita2013}. Unfortunately, numerical models have also limitations and, at the network level, analytical results are scarce, particularly for sparsely connected systems. To the best of our knowledge, the vast majority of the rigorous approaches that can be found in the literature for spiking models of neurons require some sort of \textit{diffusion approximation} (so that the membrane potential can be described approximately as a random process) \cite{Burkitt2006I,Burkitt2006II,Brunel2004}, attribute that imposes constraints on the density of links of the network  \cite{Amit1997a,Amit1997b,Brunel1999,Brunel2000}. Another mathematical approach is the so-called ``spike-train statistics'', a framework developed to analyze and fit spike patterns of population of neurons (generally driven by external stimuli or noise), by means of a characterizing probability function \cite{Cessac2008,Cessac2011}. Even though this attempt seems promising, especially to study activity patterns of real neurons, it is a proposal that is far from fully developed (there is no general strategy to derive such probability function from data, and the specific mathematical form of it is required --as well as a procedure to fit many parameters) \cite{Cessac2013}. Finally, there are reported some limited analytics based on mean-field approximations, that provide insight into very specific details of a popular family of models, known as integrate-and-fire \cite{Bressloff1999,Brunel2003,Brunel2004,Roxin2004,Olmi2010,Nicola2013,Grabska2014,Livi2016}, one of which is at the core of our analysis.
\section{Model}\label{our_model}
Our model consists of two parts. First, we generate undirected scale-free networks, i.e. random graphs whose connectivity or degree distributions are power-laws of the form, $p(k)\sim k^{-\gamma} \,\, (k_{min} > 1)$ \cite{Barabasi1999} (with their exponent in the range $2\lesssim\gamma\leq3$). To build these networks we use the configuration model \cite{Newman2003} and prescribe three additional constraints: 

\begin{enumerate}[i]
	\item The domain of the power-law degree distribution function is restricted to the values $k_{min}\leq k_{i}\leq\sqrt{N}$. 
	\item $\sum_{i}~k_{i}$ even.
	\item Self-connections are forbidden.
\end{enumerate}
Each node is thus connected to the others randomly and in accordance to these conditions, which ensures that our networks are uncorrelated \cite{Catanzaro2005}. By a straightforward integration (and normalization) one obtains the exact expression of the average degree in this case:
\begin{equation}
\langle k \rangle = \left(\frac{\gamma -1}{\gamma -2}\right)\left[\frac{k_{min}^{-\gamma+2}-N^{\frac{-\gamma+2}{2}}}{k_{min}^{-\gamma+1}-N^{\frac{-\gamma+1}{2}}} \right]\,\,\,\, (\gamma >2)\;,
\end{equation}
\\The second part of our model is the dynamics. Given a particular network of $N$ nodes, and for fixed values of the model parameters, i.e. $\tau_m$, $\tau_D$, $I_{ext}$ and $\theta$ (which from now on will be assumed constant), each node is considered as a leaky integrate-and-fire neuron that follows the equation \cite{Soula2006, Roxin2004}:

\begin{equation}
\label{eqn3.5}
\tau_{m}\frac{dV_{i}}{dt}=-V_{i}+I_{ext}+g_{i}\sum_{j,m}a_{ij}\delta\left(t-t^{(m)}_{j}-\tau_{D}\right)\,,
\end{equation}
where the time-dependent variables $V_{i}$ represent the membrane potential and $i$ the neuron index. The firing mechanism is quite simple, every time the potential of a neuron reaches or exceeds the threshold, $\theta$, it fires and its voltage drops to zero instantly. As for the parameters in \eqn~\ref{eqn3.5}, $\tau_{m}$ is the time constant (that models the decay) and the term $I_{ext}$ is typically called driving or external current, which sets the resting potential. One see that all interactions between neurons are due to the third term on the right hand side (RHS) of \eqn~\ref{eqn3.5} (usually called synaptic current). The amplitudes of the spikes are represented by $g_{i}$, $a_{ij}$ are the entries of the unweighted adjacency matrix of the network, $t^{(m)}_{j}$ the time at which neuron $j$ has fired its $m$th spike and $\tau_{D}$ is a constant pulse-delay (finite speed of the pulses). \\
Putting all these ingredients together, the dynamics goes as follows: if node $j$, for instance, has fired at $t^{(m)}_{j}$ and $a_{ij}=1$ (there is an edge between $i$ and $j$), then node $i$ receives an excitatory input of strength $g_{i}$ after $\tau_{D}$. Obviously, this dynamics is no other than a leaky I\&F (with a nonlinear synaptic currents).\\
Following Ref.~\cite{Roxin2004}, we chose $I_{ext}~<~\theta$, so neurons cannot fire by themselves (non-oscillatory regime) and it is necessary to apply some initial spike-inputs to start the activity (depending on the specific set of values of the parameters, one or several inputs are sufficient to guarantee self-sustained activity). A crucial consequence of having both pulse-delay and non-oscillator neurons is that time scale becomes discrete in the sense that there are only interactions every $\Delta t = \tau_D$. Since nothing interesting happens between consecutive interaction times, from now on we will focus exclusively on such discrete time scale.\\
As far as we know, this particular dynamics has never been studied on scale-free networks or on any other heterogeneous topology, though there are some works on random $k$--regular (amorphous) and small-world networks (most of the results, already published, are based on numerics --see references \cite{Soula2006}, \cite{Roxin2004} and \cite{Roxin2011} for more information).

\section{Basic definitions and notation}

Let $\chi_{it}$ be the firing state of neuron $i$ at time $t$ (defined to be $1$ if neuron $i$ fires at time $t$, or $0$ otherwise), then the instantaneous firing rate of the network at $t$ is defined by:
\begin{equation}
\label{eqn3.6}
{\rm Firing~Rate}(t)=\frac{1}{N}\displaystyle\sum\limits_{i=1}^{N}\chi_{it} \,,
\end{equation}
then the average firing rate of the network over a long time window ($1 \leq t \leq t_{max}$) is given by:
\begin{equation}
\label{eqn3.7}
\alpha \equiv \left\langle{\rm Firing~Rate} \right\rangle_{t}=\frac{1}{t_{max} \times N}\displaystyle\sum\limits_{t=1}^{t_{max}}\displaystyle\sum\limits_{i=1}^{N}\chi_{it}\,.
\end{equation}
The time interval between two consecutive firings of a neuron is called an inter-spike-interval (ISI, for short). For a large time window, $1\leq t \leq t_{max}$, let $N_t(i)$ be the number of ISIs of neuron $i$, and $T_{ih}$ be the duration (number of time steps $\Delta t$) of the $h$th ISI ($1\leq h \leq N_t(i)$), then ${\rm ISI_{ih}} = T_{ih} \times \Delta t $. For a system that is in a stationary regime of global self-sustained activity (i.e. transient is removed from computation), then the average ISI of neuron $i$ over a long time window is given by the expression:
\begin{equation}
\label{eqn3.8}
\left\langle{\rm ISI}(i) \right\rangle_{t} = \frac{\Delta t}{N_{t}(i)}\displaystyle\sum\limits_{h=1}^{N_{t}(i)}T_{ih} \approx  {t_{max}}/{\displaystyle\sum\limits_{t=1}^{t_{max}}\chi_{it}}\,.
\end{equation}
These definitions will be used repeatedly in \sect~\ref{neuron_theory}, and Supplementary.

\section{Numerics (homogeneous couplings)}

In order to characterize our model, in the following Sections we will analyze thoroughly several salient numerical results for homogeneous pulse strength and compare them with other findings already reported in the literature. 

\subsection{Self-sustained activity and the effects of parameters}\label{self}

\begin{figure}[htp]
	\begin{center}
		\includegraphics[scale=0.4,clip=0]{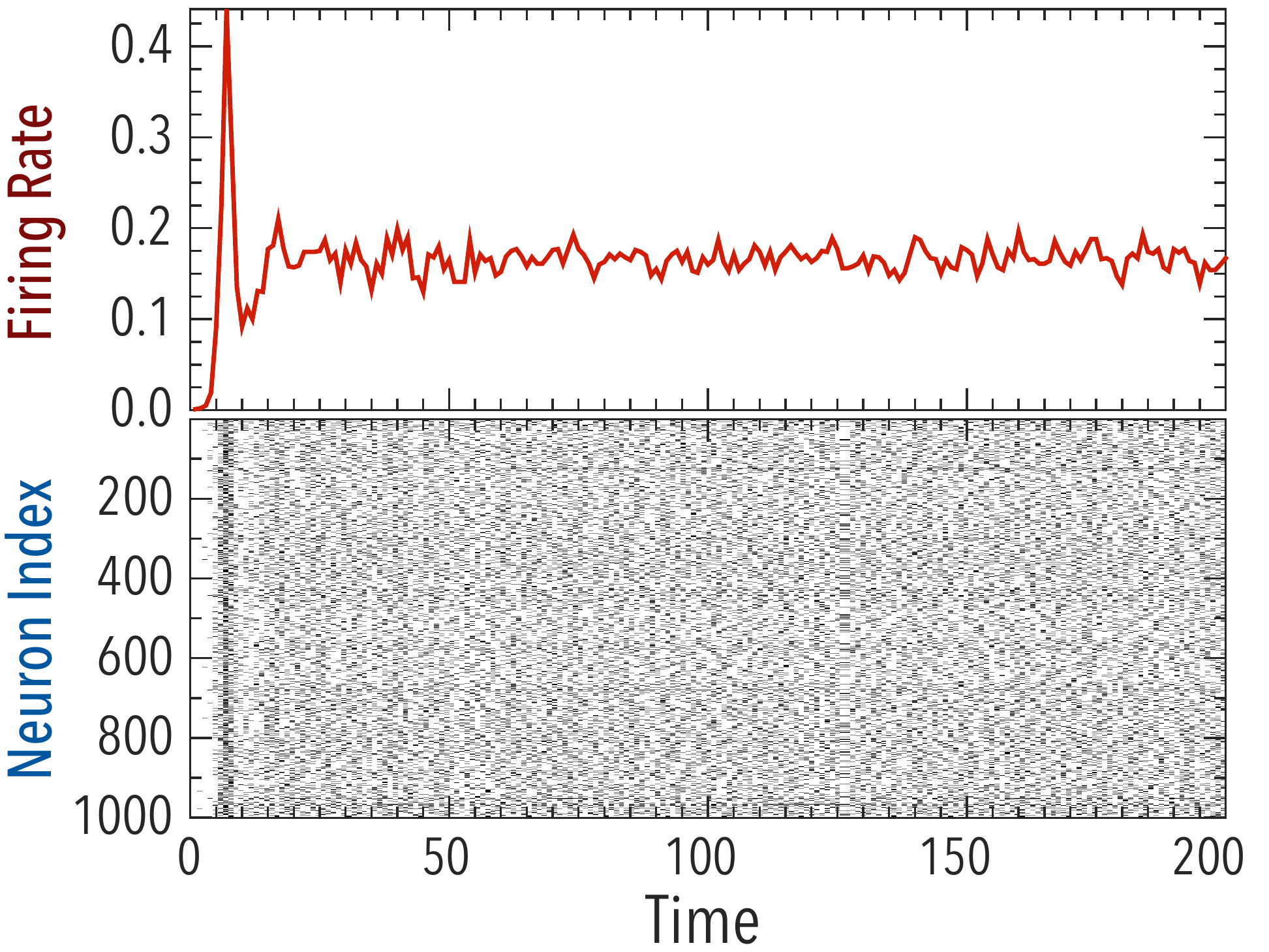}
		\caption{\label{fig3.1} Firing rate (top) and raster plot (bottom) for $N=10^{3}$, $\gamma=3.0$, $k_{min}=2$, $I_{ext}=0.85$, $g=0.2$ (constant $\forall i$), $\tau_{m}=10$, $\tau_{D}=1.0$ and $\theta=1$. Same parameter values are used in following figures unless noted otherwise.}
	\end{center}
\end{figure}

\begin{figure}[htb]
	\includegraphics[width=\columnwidth,clip=0]{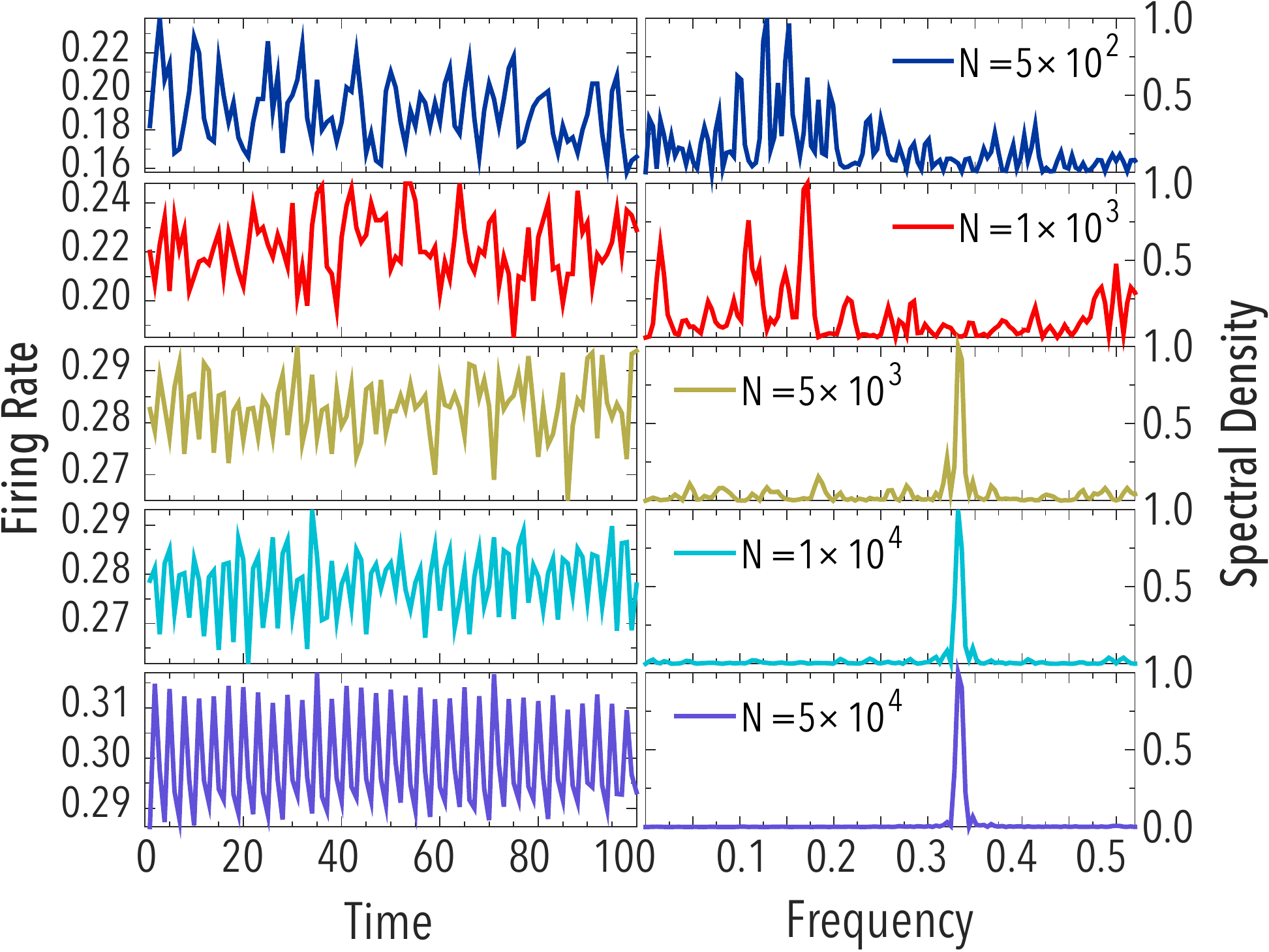}
	\caption{\label{fig3.2} Firing rate (left column) and the corresponding normalized spectral density (right column) for five network sizes. Same parameters as \fig~\ref{fig3.1}, except $N$.}
\end{figure}

\fig~\ref{fig3.1} shows an example of typical activity on SFNs: After a short transient, in which the activity spreads rapidly across the network, the global signal reaches its stable mean value (which we may call ``quiescent state'') and persists around it indefinitely. This activity seems to be self-sustained and not merely prolonged because it continues, without any additional external inputs, for a long time window $\left(t>10^{5}\right)$ with no sign of possible spontaneous failure. The underlying mechanism that explains this self-sustained activity is, as expected, related to the scale-free topology. The relative refractory period of the neurons, modeled by $\tau_{m}$, has no (or small) effect on most of the hubs because they receive many inputs (even as the reflection of their own previous spikes) at each interaction time and, therefore, they fire at every time step (leading the dynamics to a regime of high-level activity). Evidently, in \fig~\ref{fig3.1}, the global signal does not exhibit any periodicity nor significant oscillations are present, which means that there is no predominant fire frequency nor a large group of neurons that fire in-phase  (irregular firing pattern).
Taking the population firing rate in \fig~\ref{fig3.1} (left panel) as our benchmark, let us now report briefly on some of the effects of the parameters in \eqn~\ref{eqn3.5}. By varying $\tau_{m}$ or $g_{i}=g$ ($g$ constant over $i$) and leaving all the other parameters unchanged, for instance, we can alter the firing rate as follows: by reducing $\tau_{m}$ (or, equivalently, increasing $g_{i}=g$) the mean value of the global signal is increased, and we can even to change its structure (for significant variations it becomes periodic with small period --see \cite{Cessac2008}). That is to say, by changing any of those parameters in the described way, a large group of neurons saturates, because of the incoming pulses, and are forced to fire within a short time window. Thus, for some values of these two parameters, and only in terms of the global behavior of the system, having neurons that are more susceptible to small inputs is equivalent to having neurons that are being stimulated by greater impulses. Similar results can be obtained by changing the speed of the pulses (and rescaling the time axis): Fast pulses $\left(\tau_{D}<1\right)$ reduce the mean value of the firing rate (because activity is distributed more broadly over time), while slow pulses $\left(\tau_{D}>1\right)$ increase it and may even change the structure of the signal (becoming periodic in the same way as explained before).\\
As for the specific procedure to provide the initial pulse-inputs that triggers the self-sustained activity, it depends crucially on the interaction strength values $g$. Generally, for $g > \theta -I_{ext}$, one single external input applied to any neuron, at the time when $V_{i} \approx I_{ext} \; \forall i$, is sufficient to obtain self-sustained activity $(\left\langle {\rm Firing~Rate} \right\rangle_t>0)$. For the interval $(\theta-I_{ext})/k_{min} < g \lesssim \theta-I_{ext}$, several external inputs, or even the activation of all nodes, are required to star self-sustained activity, and there is no activity whatsoever for $g$ values below the lower bound of this interval (since neurons with $k=k_{min}$ cannot fire). The use of a single or $N$ initial pulse-inputs does not lead to significant differences for the average firing rate value obtained, though $N$ initial pulse-inputs might enhance (slightly) the amplitude of signal fluctuations (because more neurons fire in-phase) and also transients are usually shorter (because the activity does not have to spread across the network). Note in \fig~\ref{fig3.1} (left panel) that when a single fire is used to start the dynamics, a sharp peak appears during the transient (about $\sim50\%$ of neurons fire at once) and then the signal decreases to the quiescent state. This feature suggests that a global (simultaneous) activation of neurons to initiate activity is a strategy that might not affect significantly the activity observed after the transient. However, for $g = [\theta-(1-e^{-\frac{\Delta t}{\tau_{m}}})I_{ext}]/k_{min}$, network activity is already saturated $(\left\langle {\rm Firing~Rate} \right\rangle_t=1)$ for $N$ initial pulse inputs while it might not be yet for a single one (but close). \\
As shown in \fig~\ref{fig3.2}, there is a size effect related to the structure of the firing rate and its mean value. Again, for the smallest networks illustrated in the left panels of \fig~\ref{fig3.2} ($N \leq 10^{3}$) no predominant fire frequency in the global signal is observed, thus we say that the firing rate has no periodicity. Whereas for the largest networks considered $\left(N \gtrsim 10^{4}\right)$,  without changing any parameter of the dynamics, signal structure emerges (reported in the left column of \fig~\ref{fig3.2}). To corroborate the presence of such periodicities we computed normalized spectral densities for each firing rate (displayed in the right column of \fig~\ref{fig3.2}), estimated using the discrete Fourier transform method,
\begin{equation*}
Y(k)=\displaystyle\sum\limits_{j=1}^{n} v(j)\omega_{n}^{(j-1)(k-1)} \,, {\mbox{ where }}\,\omega_{n}=e^{-2\pi i/n} \,.
\end{equation*}

\begin{figure}[htb]
	\begin{center}
		\includegraphics[width=\columnwidth,clip=0]{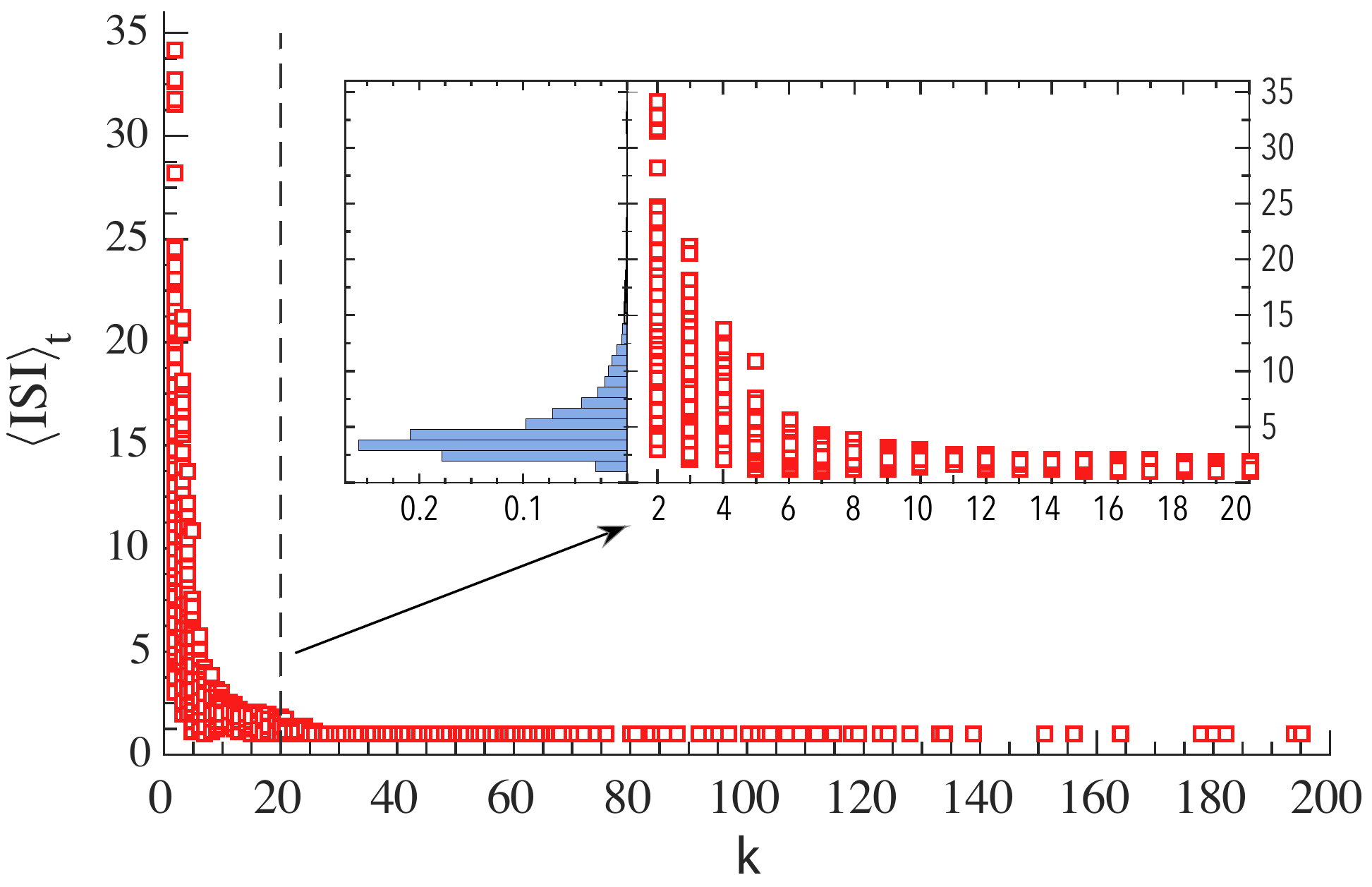}
		\caption{\label{fig3.3} Main: Scatter plot of average ISI. The averages were computed over $10^{5}$ time steps. Same parameters as \fig~\ref{fig3.1}, except $N=5\times10^{4}$. Inset: Scatter plot of average ISI for $k\leq20$ (right) and its corresponding histogram (left).}
	\end{center}
\end{figure}

Obviously, in \fig~\ref{fig3.2}, there can be seen multiple peaks for the smallest networks whereas for the largest networks just a single peak. To explain this finding we have to go back to our network model. Our networks are generated following three prescriptions, one of them being $k_{max}\sim\sqrt{N}$. So the larger the network, the greater the amount of hubs and their degrees. Figure~\ref{fig3.3} (main) depicts $\left\langle\mbox{ISI}(i)\right\rangle_{t}$ for all neurons of the network of size $N=5 \times 10^{4}$, and any of them with $22 \leq k \lesssim 200$ has an average ISI equals to 1 (i.e. complete saturation, they fire at every time step). Consequently, a relevant set of saturated hubs arises for large networks, that do not exist for small ones, which drives the activity of small degree neurons to fire every few time steps and also increases the average value of the global signal. An additional confirmation can be obtained from the left panel of the inset in \fig~\ref{fig3.3}, the histogram of mean ISI, which reveals that more than one-third of the total neurons (highest bin) have $\left\langle\mbox{ISI}(i)\right\rangle_{t}=3$, value that corresponds to the frequency detected by spectral analysis. This means that the structure of the signal is formed by adding individual signals of neurons within the smallest connectivities groups ($k=2,~3$, since these are the most abundant). The other signals that compose the global firing rate, neurons having $k \gtrsim 4$, just add to the mean value and not to the periodicity. Periodic oscillations of neural populations have already been reported for different models of neurons on a few complex topologies \cite{Araujo2006,McGraw2011,Roxin2004}, but specifically small period oscillations for a leaky I\&F model, to our knowledge, only by reference \cite{Cessac2008}.\\
A most significant observation from the previous results, and from the extensive numerical simulations, is that the stationary average firing rate does not increase continuously from zero when increasing $g$, though it does in an abrupt way; in other words, there is a threshold value $g_c$ (script ``$c$'' stands for critical) at which self-sustained activity starts with an average firing rate well above zero. Another relevant remark on the average firing rate with respect to the coupling is that it is a monotonically increasing function of $g$, but not a strictly increasing one (so for some values of $g$ the average firing rate might not change).\\
Finally, the salient observation, already mentioned, that once in the self-sustained activity regime (for $g > g_c$), there is a set of extremely active neurons for which $\chi_{it} =1$ for all time $t$, leads to the notion of ``saturation degree'' $k^s$ (that decreases with $g$) such that for all neurons $i$ with degree $k_i \geq k^s$, their ${\rm ISI}(i)=1$. Again, this core of relatively high degree/fast firing neurons is responsible for the maintenance of global self-sustained activity in SFNs and, for large networks, is also responsible for the emergence of periodic oscillations with a higher mean value of the firing rate. We will use some of these definitions and quantities in \sect \ref{neuron_theory}.

\subsection{Bistability and Signal Control}\label{control}

\begin{figure}[htb]
	\includegraphics[scale=0.4,clip=0]{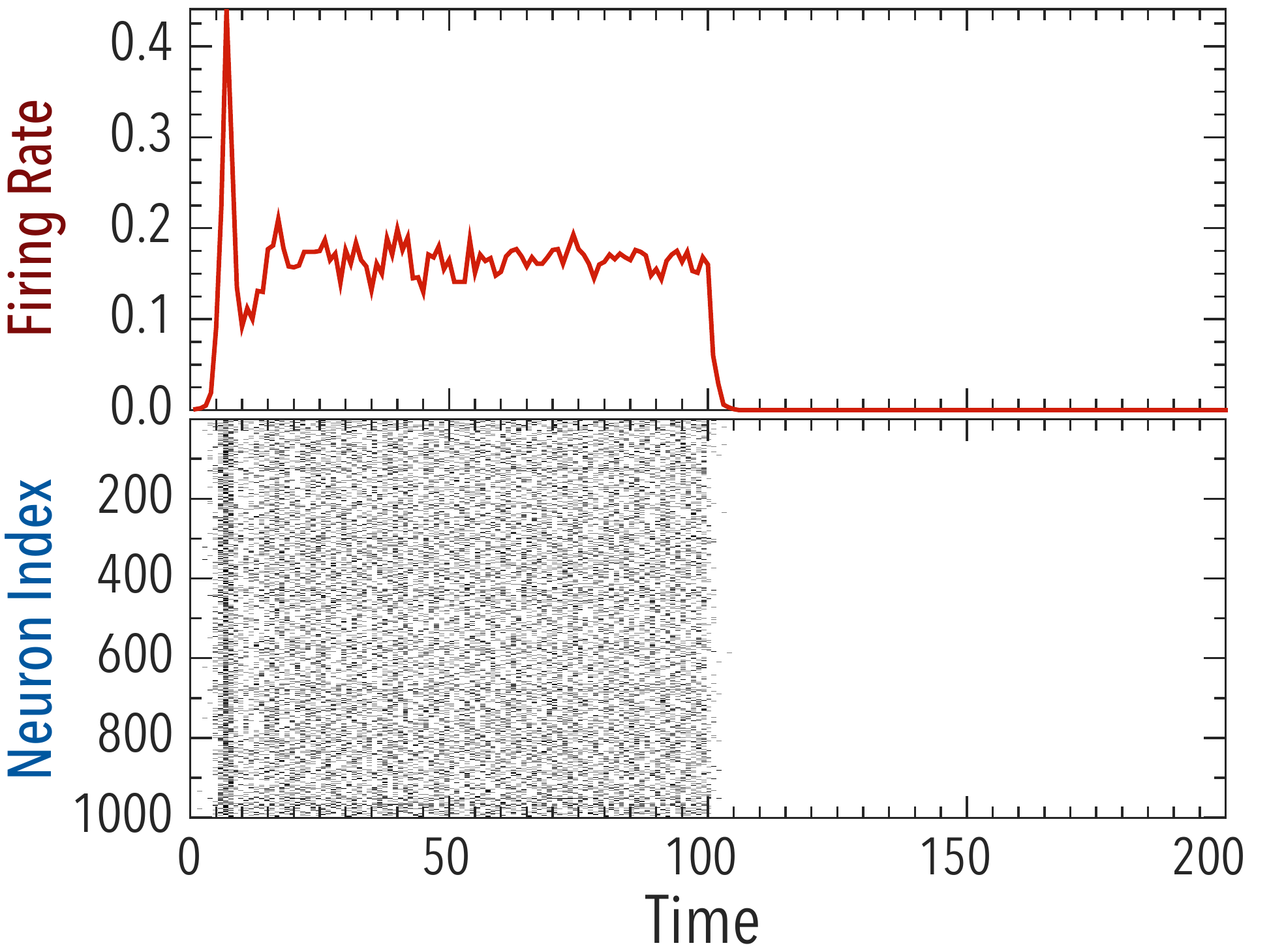}
	\caption{\label{fig3.6} Firing rate (top) and raster plot (bottom). Bistable behavior of the network. At $t=100$ an inhibitory global signal ($60\%$of neurons) was applied to turn off the dynamics. Same parameters as \fig~\ref{fig3.1}.}
\end{figure}

\begin{figure}[htb]
	\includegraphics[scale=0.4,clip=0]{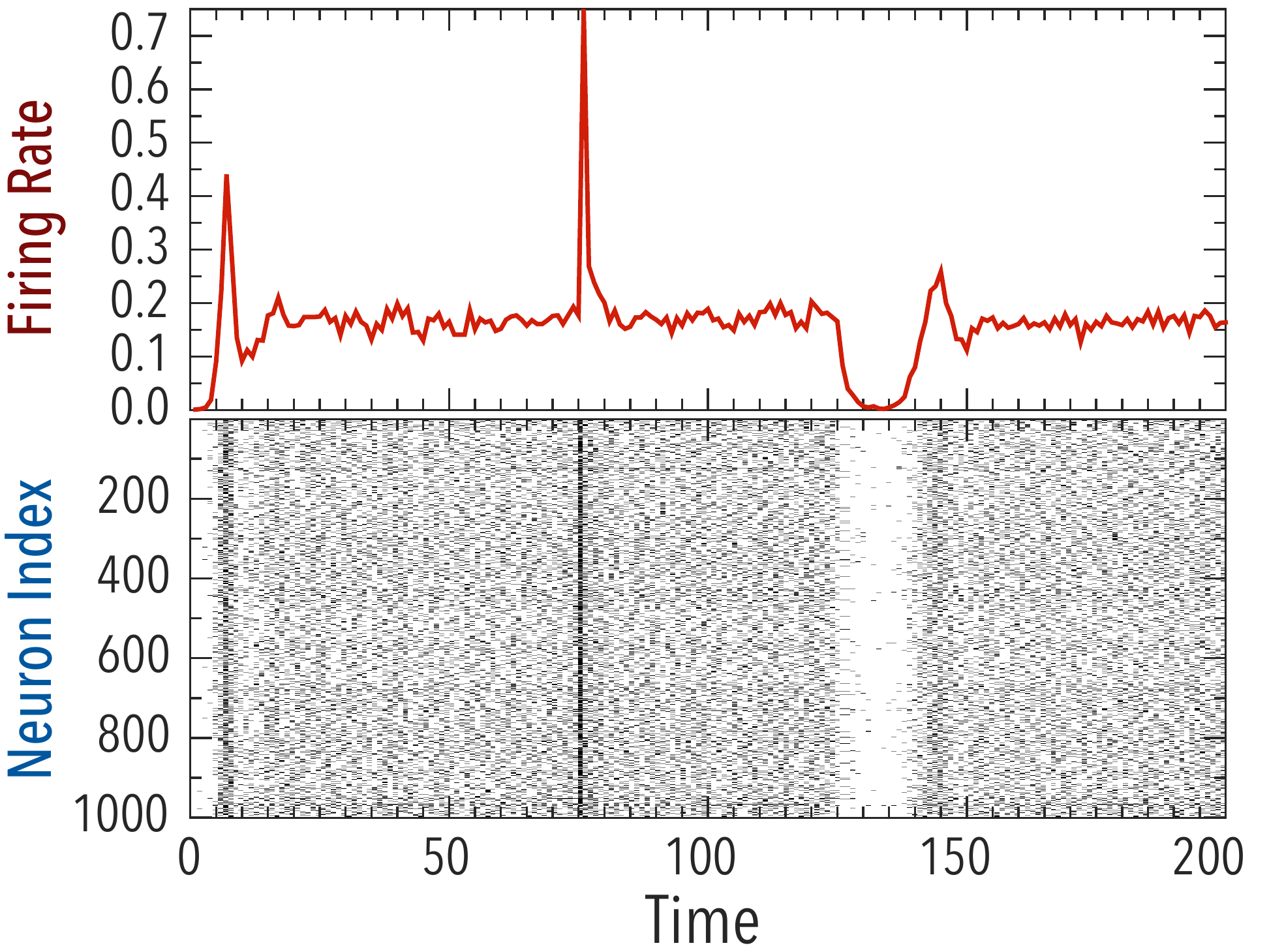} 
	\caption{\label{fig3.4} Firing rate (top) and raster plot (bottom). Same parameter as \fig~\ref{fig3.1}.}
\end{figure}

\begin{figure}[htb]
	\includegraphics[scale=0.4,clip=0]{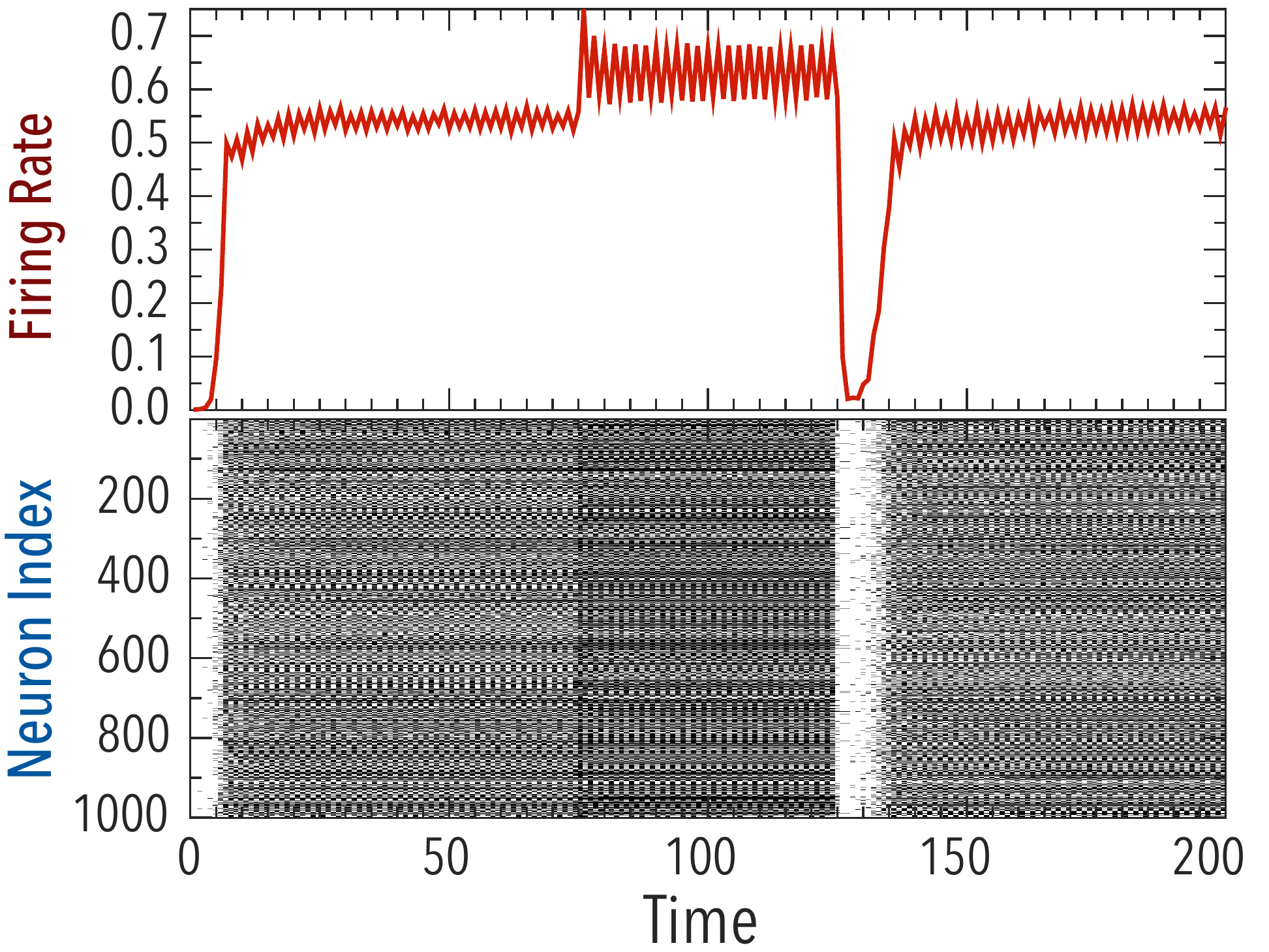} 
	\caption{\label{fig3.5} Firing rate (top) and raster plot (bottom) for $g=0.35$. It is possible to change the amplitude of the oscillations of the global signal, but its structure (periodicity and shape) remains the same.}
\end{figure}

It is possible to perturb the ongoing dynamics at any point of the quiescent state by activating or deactivating randomly chosen neurons (i.e. setting to $1$ or $0$ the firing states $\chi_{it}$, respectively, and resetting to zero their membrane potentials $V_{i}$). This procedure is biologically feasible and it can be understood as a result of the interaction between the scale-free neural network and, for instance, a pacemaker or even a regulatory neural network.  
In the next paragraphs we will describe some of the most representative responses of the system that we have found when using this approach:
{\bf(i) Bistability:} For the smallest networks considered here ($N \lesssim10^{3}$)  an inhibitory signal applied to $\gtrsim 60\%$ of  all neurons is enough to extinguish the activity permanently (see \fig~\ref{fig3.6}), but for large networks ($N \sim 10^{5}$, not shown) only consecutive inhibitory global-pulses (at least two) are capable of turning ``off'' the dynamics. The activity can be triggered again, in the usual way, once the membrane potentials of all neurons have reached the resting value $I_{ext}$. Hence, the property called bistability, which is precisely the capability to switch the neural activity between ``on'' and ``off'' (resting) states, is also present in our model \cite{Roxin2004}. Of course, all these remarks are valid for the parameters used and as soon as they are modified, for instance, reducing the activity level, then less inhibition might be needed in order to cease all activity. An $ad~hoc$ mechanism to turn back ``on'' the network can be easily introduced using noise (allowing neurons to fire randomly) but will not consider this approach here (see reference \cite{Roxin2004} for more details).
{\bf(ii) No signal response:} \fig~\ref{fig3.4} shows the same activity pattern depicted in \fig~\ref{fig3.1}, but in this case with a large  excitatory stimulus at $t=75$ ($75\%$ of the total neurons) and an inhibitory perturbation at $t=125$ ($\lesssim 60\%$ of the total neurons). The first aspect to notice from it, is that for these parameters the global-pulses have no persistent effect on the global firing signal, because neither through excitatory nor inhibitory means was possible to modify the mean value or the structure of the firing rate after the perturbation' transient. The second aspect to acknowledge is that the time needed to return to the quiescent state (recovery time) for inhibitory perturbation is much longer than the recovery time for excitatory stimulus, which is reasonable because the activity has to spread across the network again.
{\bf(iii) Activity enhancement (and hindrance):} \fig~\ref{fig3.5} illustrates the dynamics for a stronger coupling strength. The firing rate is clearly periodic in this case (by saturation effect, described in the previous Section), and at $t=75$ the excitatory stimulus enhances the amplitude of the oscillation. As already explained, this enhancement is due to the fact that, after the excitatory stimulus is applied, there are more neurons (with ISI close to the period of the global signal) firing in-phase. To show that it is possible to recover the initial amplitude of the oscillations, at $t=125$ an inhibitory perturbation is applied to $90\%$ of the neurons and obviously it is sufficient to reduce the oscillations. Although this neural stimulus acts on randomly chosen neurons, it might be considered as a globally reversible process when using certain parameters. Note that the recovery time to reach the quiescent state is shorter because of the high-level activity.\\

\subsection{Activity Failure and Dynamical Resilience}

\begin{figure*}[htb]
	\centering
	\begin{tabular}{lc}
		\includegraphics[scale=0.435]{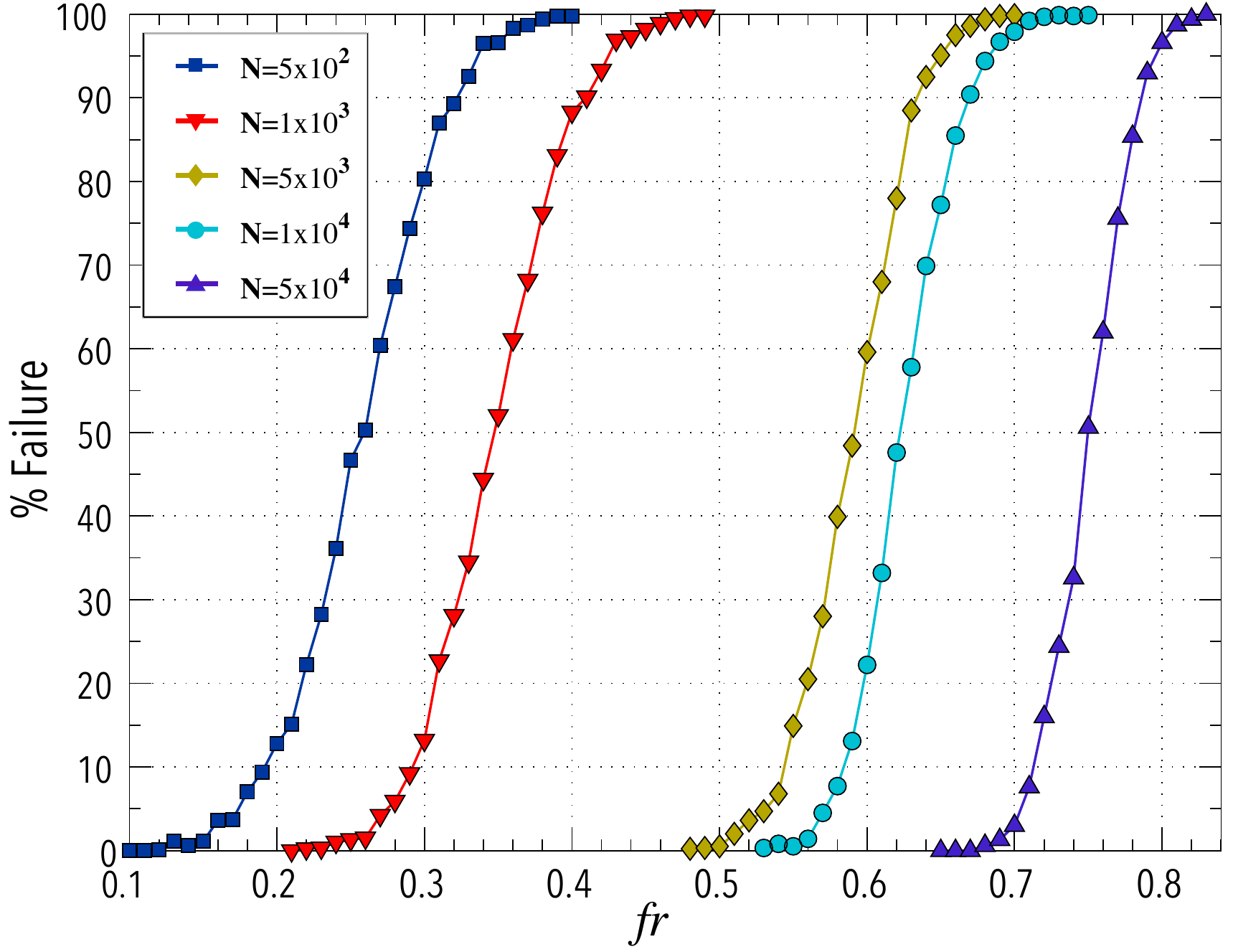}&
		\includegraphics[scale=0.405]{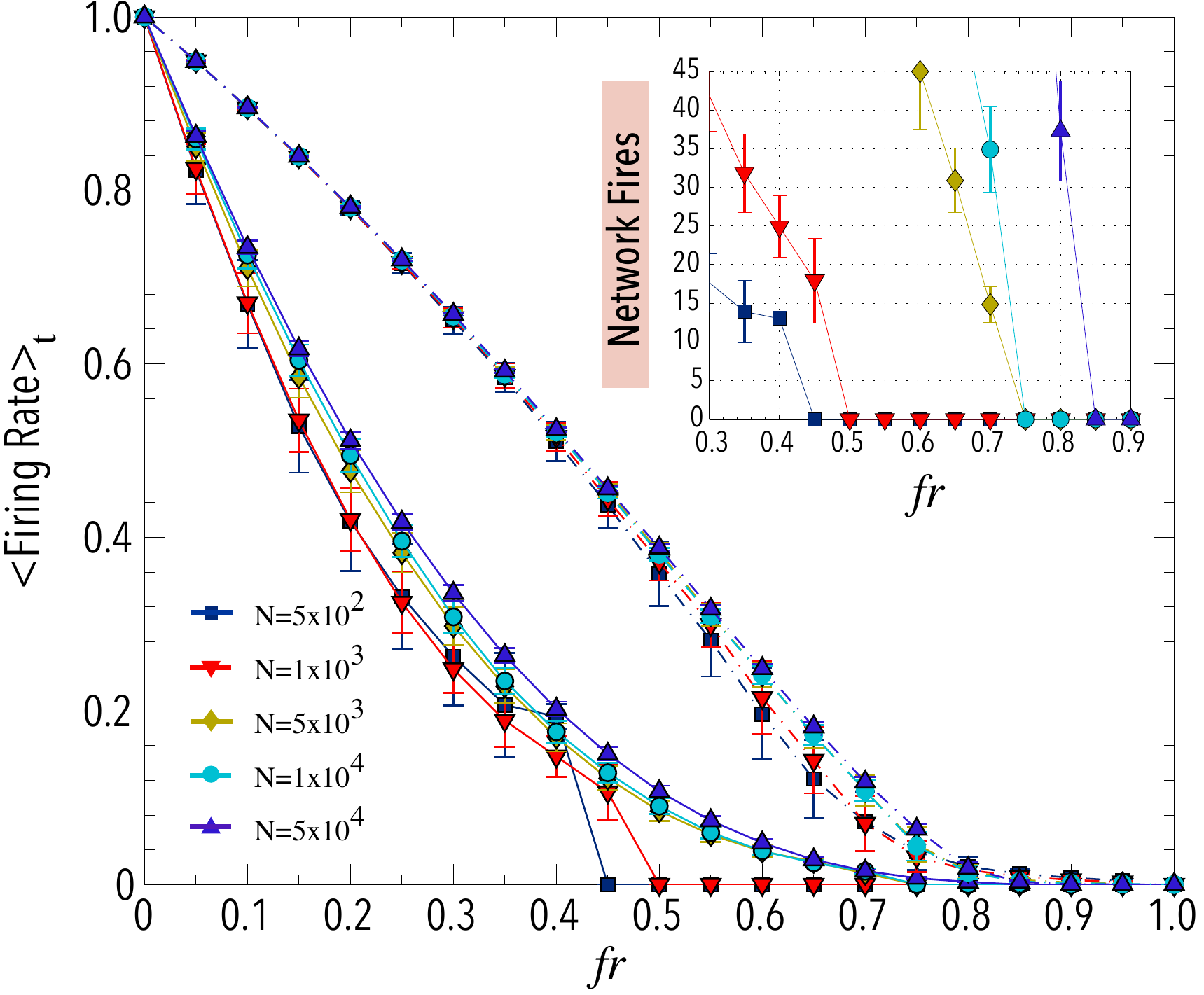}
	\end{tabular}
	\caption{\label{fig3.7} Failure activity probability as a function of the fraction of randomly removed nodes (left panel). Main (right panel): average value of the firing rate over time (solid curves, all values normalized by $\left\langle {\rm Firing~Rate} \right\rangle_t$ at $fr=0$) as a function of the fraction of randomly removed nodes, and the corresponding size of the giant component (dashed curves) of the percolated network. Inset (right panel): Total fires as a function of the fraction of randomly removed nodes. Same parameters as the ones used in \fig~\ref{fig3.1}.}
\end{figure*}

To investigate the activity failure as a result of the damaging process of the topology of the network, we performed an uniformly random removal of nodes on several SFNs \cite{Newman2003} and then we executed the dynamics on the giant component of the resulting network. \fig~\ref{fig3.7} (left panel) exhibits the failure probability as a function of the removed fraction of nodes ($fr$), but taking into account as failure only those realizations for which all activity ceased before $t=t_{transient}+200$ time steps (each point on the curves is an average over $10^{3}$ realizations of the random deletion of nodes). It is possible, of course, that some realizations assessed as successful using the criterion mentioned above were just prolonged activity instead of self-sustained, but such criterion turned out to be quite reliable --in most cases-- because activity failure in SFNs generally occurs within a small time-window (perhaps another consequence of being ultra-small \cite{Cohen2003}). Note that the steepness of failure probability curves in the transition regime increases with the size of the network (as it should be).
Another aspect to consider here, perhaps one that describes better what occurs with the dynamics in the damaging process, is illustrated in the right panel of \fig~\ref{fig3.7}, where the average firing rate (of successful realizations) is shown as a function of the fraction of deleted nodes (solid curves, each point is the average over $200$ non-null realizations, i.e. for $\left\langle {\rm Firing~Rate} \right\rangle_t>0$). The curves of the size of the giant component (again, average over $200$) are also shown for comparison in the right panel of \fig~\ref{fig3.7} (dashed curves). Clearly, the average firing rate is a decreasing function in all cases when the removal is performed, but for the largest networks complete dynamical failure only occurs close to the critical value of the fraction at which the network has percolated completely ($fr\approx 0.8$). In this sense, activity on SFNs is quite resilient because, in spite of crucial changes in network' structure, it finds its way to persist until the complete disintegration of the topology is almost achieved. The inset in the right panel of \fig~\ref{fig3.7} confirms this results, but also shows an unexpected outcome of the last stages of the damaging process: Strikingly, the total number of networks fires (unnormalized) just before complete failure for all network sizes are of the same order of magnitude, which means that all SFNs are capable of sustaining activity until certain number of nodes, regardless of the original size of the network.

\section{Analytical estimations}\label{neuron_theory}

In an effort to provide some rationale for these salient observed features, and on the basis of some simplifying approximations, we develop in this Section some analytics whose predictions agree qualitatively, but also quantitatively in some respects, with observations on the stationary regime of self-sustained activity.\\
Again, let us begin by assuming that the system is in a stationary regime of global self-sustained activity on SFNs. In this framework, a recurrence relation on $V_{i}(t)$ can be derived by straightforward integration of \eqn~\ref{eqn3.5}:
\begin{equation*}
V_i\left(t\right) =  V_i\left(t-\Delta t\right) e^{-\frac{\Delta t}{\tau_{m}}} + (1-e^{-\frac{\Delta t}{\tau_{m}}})I_{ext} + g b_i\left(t\right)\;, 
\end{equation*}
where $b_i(t)$ is the number of pulse-inputs that neuron $i$ receives at time $t$. Then the firing condition at the end of the $h$th ISI is expressed as,
\begin{equation}
\label{eqn3.9}
\displaystyle\sum\limits_{n=0}^{T_{ih}-1}\left[\left(1-e^{-\frac{\Delta t}{\tau_{m}}}\right)I_{ext}+g\,b_{(T_{ih}-n)}\right]
e^{-\frac{n \Delta t}{\tau_{m}}} \gtrsim \theta
\end{equation}
where we use a simplified notation for the number of pulse-inputs received at each time step, within the $h$th ISI, by neuron $i$: $b_{1}$ are the inputs received by neuron $i$ at $t=\Delta t\sum_{j=1}^{h-1}T_{ij}$ (the instant when the $h$th ISI begins), $b_{2}$ are the inputs received by neuron $i$ at $t=\Delta t\sum_{j=1}^{h-1}T_{ij}+\Delta t$, ... , $b_{T_{ih}}$ are the inputs received by neuron $i$ at $t=\Delta t\sum_{j=1}^{h}T_{ij}$ (the instant when the $h$th ISI ends). 

\begin{figure}[htp]
	\includegraphics[width=\columnwidth,clip=0]{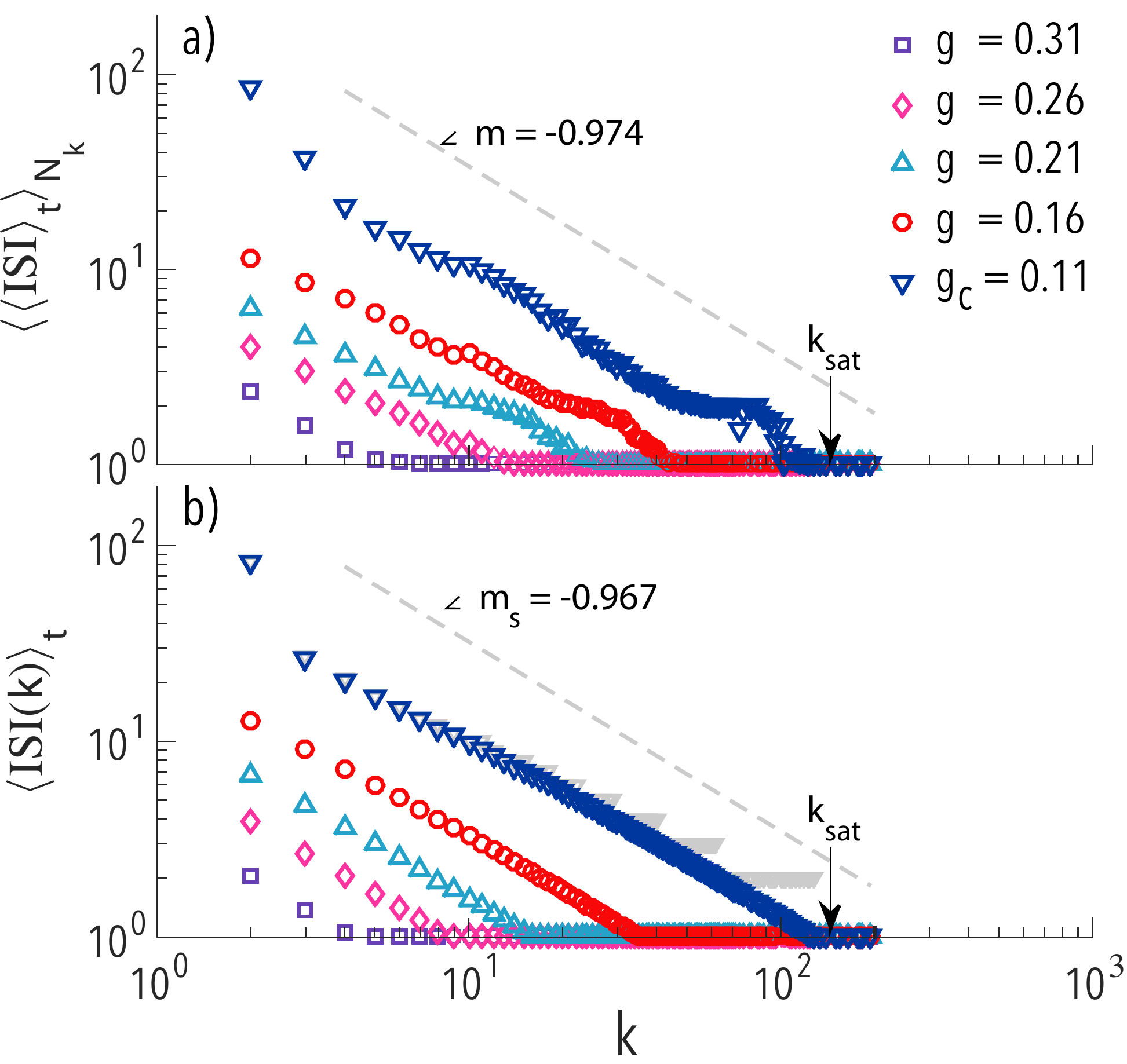}
	\caption{\label{fig3.8} ISI heterogeneity. (a) Numerical result: average ISI (computed over $10^{5}$ time steps) vs. connectivity and (b) Prediction of \eqn~\ref{eqn3.11}. The size of the network is $N=5\times10^{4}$ and the critical coupling for this network is $g_c = 0.11$ (blue downward triangles). Note that some numerical curves display what we call ``step effect'' as a consequence of the discrete time scale imposed by pulse delays. In panel (b), only for this critical value $g_c$, gray triangles represent $\ceil{T(k)}\times \Delta t~~\forall k$ and are displayed for comparison. From \eqn~\ref{eqn3.15} we estimated $k_{sat} \approx 130$ and is shown to compare with the numerical result, $k_{sat}=128$. Finally, dashed gray lines are exhibited as a measure of agreement --they do not mean power-law behavior--, the relative error between their slopes is $E_{\%}~\approx~0.7\%$.}
\end{figure}

After averaging equation \eqn~\ref{eqn3.9} over all ISIs of neurons having degree $k$, as well as using a few simplifying (mean-field-like) assumptions one obtains (see Supplementary), for the average $T(k)$ of $T_{ih}$ in the $k$-class:
\begin{equation}
\label{eqn3.10}
T(k) \approx \frac{\tau_{m}}{\Delta t}\; \ln\left[\frac{\left(1-e^{-\frac{\Delta t}{\tau_{m}}}
	\right)I_{ext}+g\alpha k}{\left(1-e^{-\frac{\Delta t}{\tau_{m}}}
	\right)\left(I_{ext}-\theta \right)+g\alpha k}\right]\;,
\end{equation}
where $\alpha$ is the average firing rate, given by \eqn~\ref{eqn3.7}. The quantity $T(k)$ provides the average ISI of neurons of degree $k$ as:
\begin{equation}
\label{eqn3.11}
\left\langle{\rm ISI}(k) \right\rangle_{t}=\begin{cases}
\ceil{T(k)}\times\Delta t &{\mbox{ for }}\; T(k) \leq 1\\
T(k)\times\Delta t &{\mbox{ for }}\; T(k) > 1\;,
\end{cases}
\end{equation}
where $\ceil{x}$ represents the ceiling function. As expected, $T(k)$ is a monotone decreasing function of $k$. Using as an input the numerical results of the stationary average firing rate $\alpha$ in \eqn~\ref{eqn3.10}, for different values of the coupling $g$, one can obtain estimations of $\left\langle{\rm ISI}(k) \right\rangle_{t}$ and compare them with the numerical results (see \fig~\ref{fig3.8}). Clearly, \eqn~\ref{eqn3.11} provides a good overall description of the average behavior of each $k$-class.

\eqns~\ref{eqn3.10} and \ref{eqn3.11} are useful to derive other properties of the model. For instance, the value $k= k^s$ can thus be estimated for any value of $g$ (provided $\alpha$), as $\left\langle{\rm ISI}(k^{s}) \right\rangle_{t}=1$ (the script ``$s$'' stands for saturation, as mentioned in the last paragraph of \sect \ref{self}): 
\begin{equation}
\label{eqn3.12}
k^{s} \equiv \frac{1}{g\alpha }\left[\theta-\left(1-e^{-\frac{\Delta t}{\tau_{m}}}\right)I_{ext}\right] \;,
\end{equation}
so that all neurons with degree $k\geq k^s$ are saturated. Note that the saturation degree value is a decreasing function of $g$ (for $\alpha$ is naturally a non-decreasing function of $g$).
As $T(k)$ must be a positive quantity, again from \eqn~\ref{eqn3.10} one has that, for all $g$ and $k$, 
\begin{equation}
\label{eqn3.13}
\alpha > \frac{1}{g k}\left(1-e^{-\frac{\Delta t}{\tau_{m}}}\right)\left(\theta-I_{ext} \right) \;.
\end{equation}
The RHS of this inequality is maximum at $k=k_{min}$, and therefore, as $\alpha$ is a non-decreasing function of the interaction strength $g$, one concludes that there is a lower bound $g_c$ below which there is no global self-sustained activity, and that $\alpha$ jumps, at $g=g_c$, from zero to the finite value
\begin{equation}
\label{eqn3.14}
\alpha_{c} \equiv \alpha(g_{c}) = \frac{\left(1-e^{-\frac{\Delta t}{\tau_{m}}}\right)\left(\theta -I_{ext}\right)}{(g_{c} \times k_{min})}\;.
\end{equation}
Obviously $\alpha_{c}$ is just a lower bound for the average firing rate and, since SFNs are ultra-small \cite{Cohen2003}, it just takes a few integration steps of the dynamics in order to ascertain whether there is activity or not after the transient. So in this way, from numerical simulations, we can determine easily the minimum value $g=g_{c}$ at which any network exhibits self-sustained activity, and thus we can estimate $\alpha_{c}$ from \eqn~\ref{eqn3.14}. 
As an example, the estimation of $\alpha_c$ for the network used in \fig~\ref{fig3.8} is $\alpha_{c}=0.0648$ and, from numerics, $\langle{\rm Firing~Rate}\rangle_{t}=0.0650$ at the same coupling strength, so strikingly there is only a small relative error of $E_{\%}~\approx~0.3\%$ between numerics and analytics. The fact that the expression for $\alpha_c$ provides, in this case,  a good estimation of the average firing rate obtained from simulations means that this dynamics on SFNs is quite resilient, because it holds until a small value of the coupling, at which most of the $k_{min}$-class neurons are about to fail completely their activity. 
From \eqns~\ref{eqn3.12} and \ref{eqn3.14} we can calculate the saturation degree ($k^{s}$) at the critical interaction strength ($g_c$), that we symbolize as $k_{sat}$, 
\begin{equation}
\label{eqn3.15}
k_{sat}~\equiv~k^{s}(g_c) = \left[\frac{\theta -\left(1-e^{-\frac{\Delta t}{\tau_{m}}}\right)I_{ext}}{\left(1-e^{-\frac{\Delta t}{\tau_{m}}}\right)\left(\theta-I_{ext}\right)}\right]k_{min}\;,
\end{equation}
which turns out to be completely determined by the (isolated) neuron parameters and the smallest degree $k_{min}$ of the network (then its computation does not require any input from numerics). This result for the saturated core at the critical coupling is quite surprising because it was expected to find that the larger the network, the larger $k^{s}$ (since $k_{max}\sim\sqrt{N}$), but now it is clear that $k_{sat}$ does neither depend on the size of the network nor on the specific value of $g_{c}$. See caption of \fig~\ref{fig3.8} for an example of the quantitative agreement obtained, prediction that we have successfully tested for very large SFNs (up to $N = 10^5$). 

Another quantity that is also completely determined by the single neuron parameters is the slope, $m_{s}$, of the curve $\left\langle{\rm ISI}(k) \right\rangle_{t}$ at $k=k^{s}$ for $g=g_c$ in a log-log plot (see dashed line in \fig~\ref{fig3.8}):
\begin{equation}
\label{eqn3.16}
m_{s}= \frac{\tau_{m}}{\Delta t \theta}\left(1-e^{\frac{\Delta t}{\tau_{m}}}\right)\left[\theta-\left(1-e^{-\frac{\Delta t}{\tau_{m}}}\right)I_{ext}\right]\;,
\end{equation}
which also turns out to be independent of the critical coupling value $g_c$. This slope might be useful because there is an almost-linear behavior (in log-log plot) for a wide range of $k$, so it can be used a measure of agreement between numerics and analytics.  

From all these results it is indisputable that, even without any further development, \eqn~\ref{eqn3.10} provides useful insight into this dynamics and also allows us to predict a few significant consequences of our model from a single input from numerics. Nevertheless, we have derived an approximate relation (see Supplementary) for the average firing rate $\alpha$ and the average ISI of the $k$-class:
\begin{equation}
\label{eqn3.17}
\alpha = \left\langle{\rm Firing~Rate} \right\rangle_{t} \approx \displaystyle\sum\limits_{k=k_{min}}^{k_{max}}\frac{p(k)}{\left\langle{\rm ISI}(k) \right\rangle_{t}}\;,
\end{equation}
which is only approximate due to the finite value of $t_{max}$. \eqns~\ref{eqn3.10}, \ref{eqn3.11} and \ref{eqn3.17} constitute a set of coupled non linear equations which can be solved to estimate $\left\langle{\rm ISI}(k) \right\rangle_{t}$ and $\alpha$ without the recourse to simulations. 

\subsection*{Accuracy of our approach}

In order to examine how accurate the assumptions and approximations made in deriving \eqn~\ref{eqn3.17} are, we introduce the following implicit equation (that can be solved without performing any computer simulation of the dynamics):

\begin{equation}
\label{eqn3.19}
f(\alpha)= \alpha - \displaystyle\sum\limits_{k=k_{min}}^{k_{max}}\frac{p(k)}{\left\langle{\rm ISI}(k) \right\rangle_{t}}=0
\end{equation}
\begin{figure*}[htb]
	\centering
	\begin{tabular}{lc}
		
		\includegraphics[height=50mm]{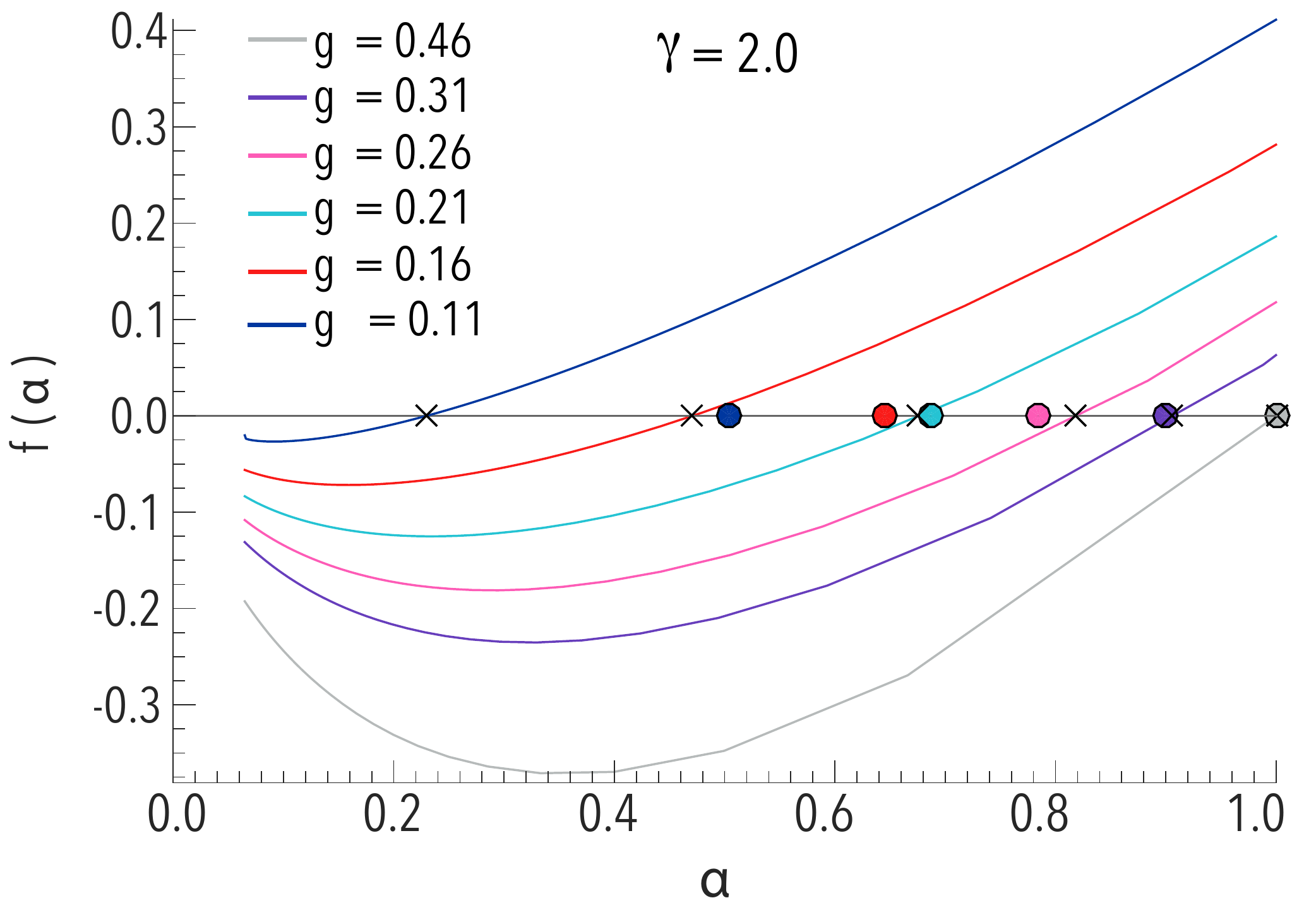}&
		\includegraphics[height=50mm]{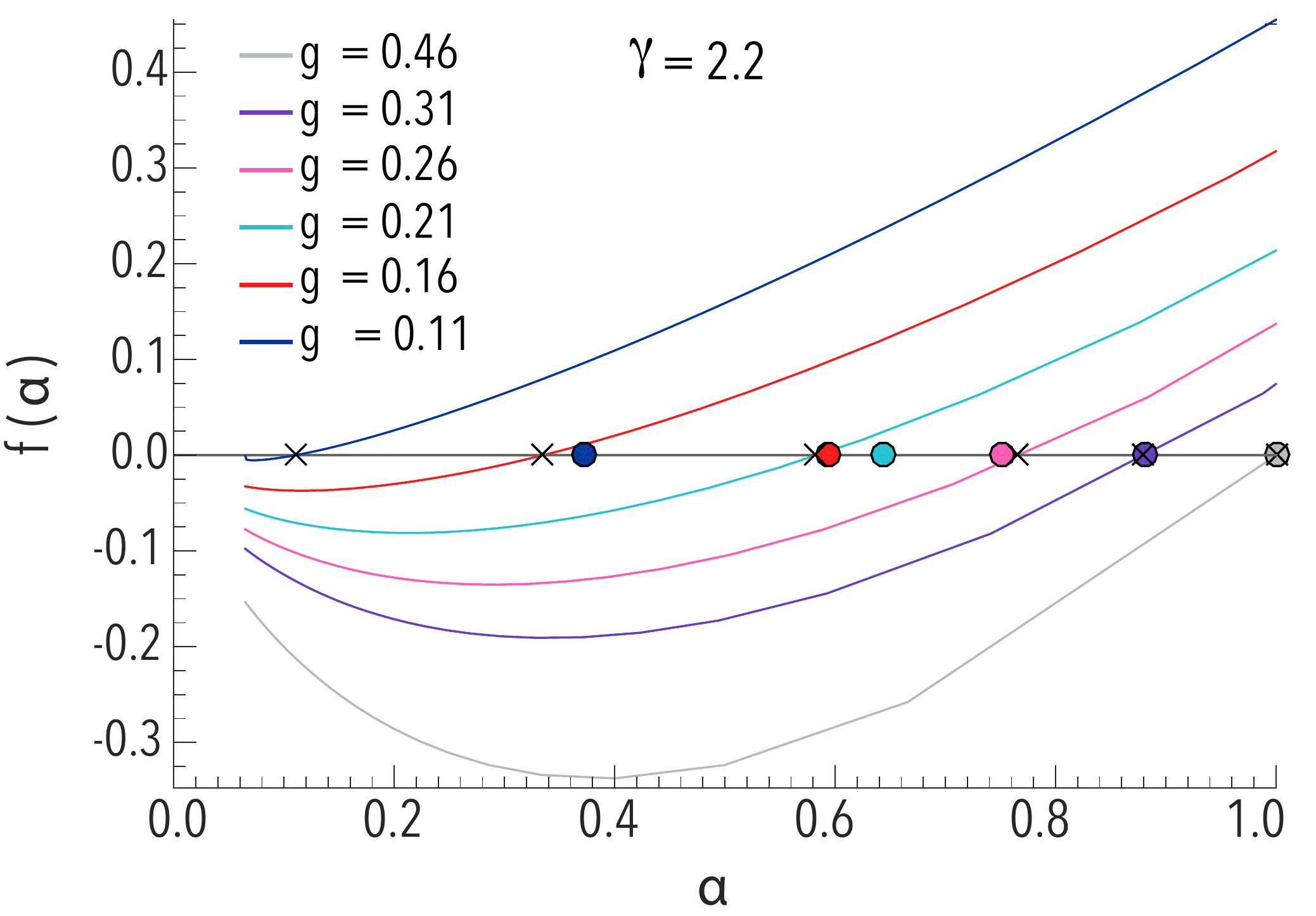}\\
		\includegraphics[height=50mm]{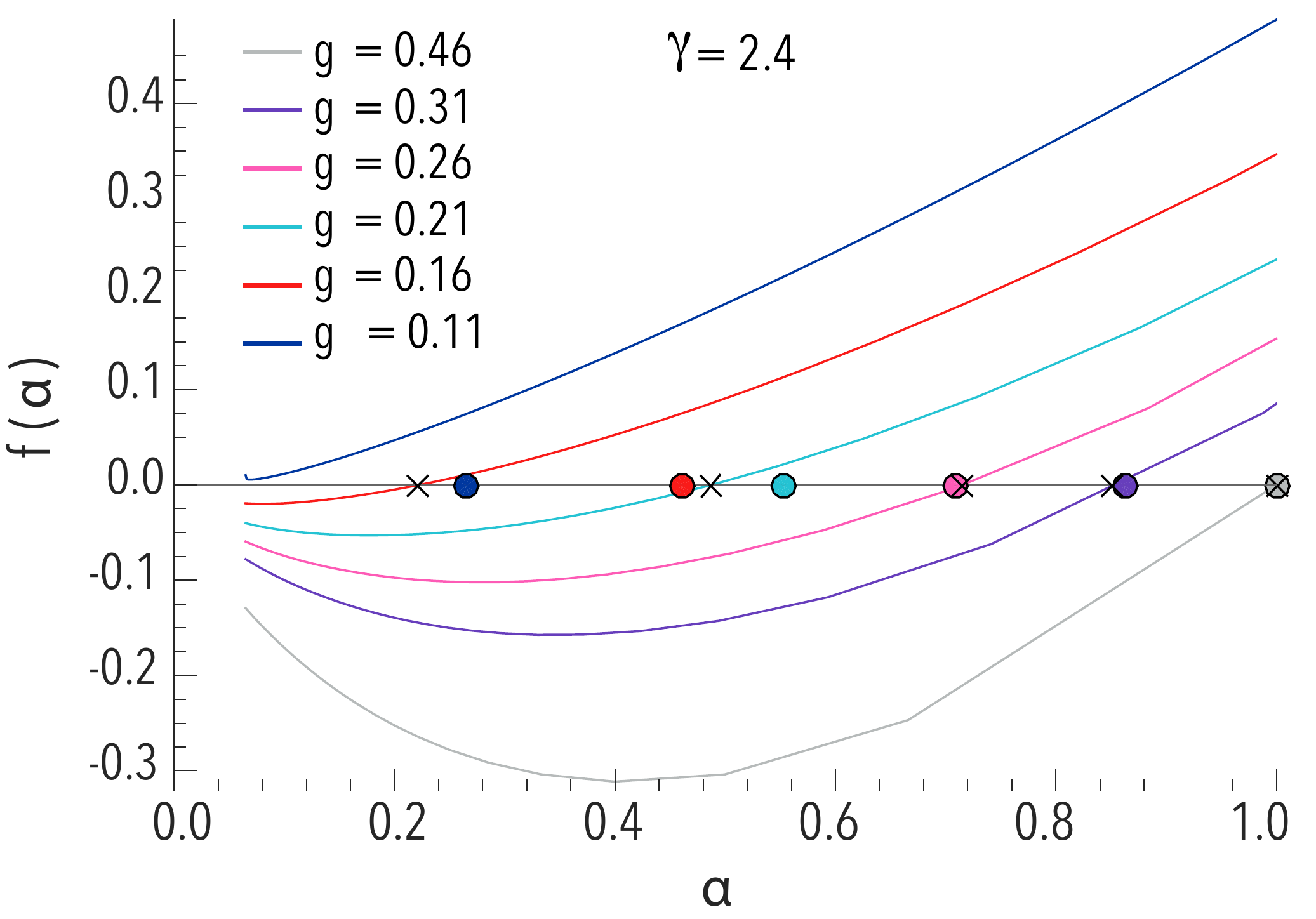}&
		\includegraphics[height=50mm]{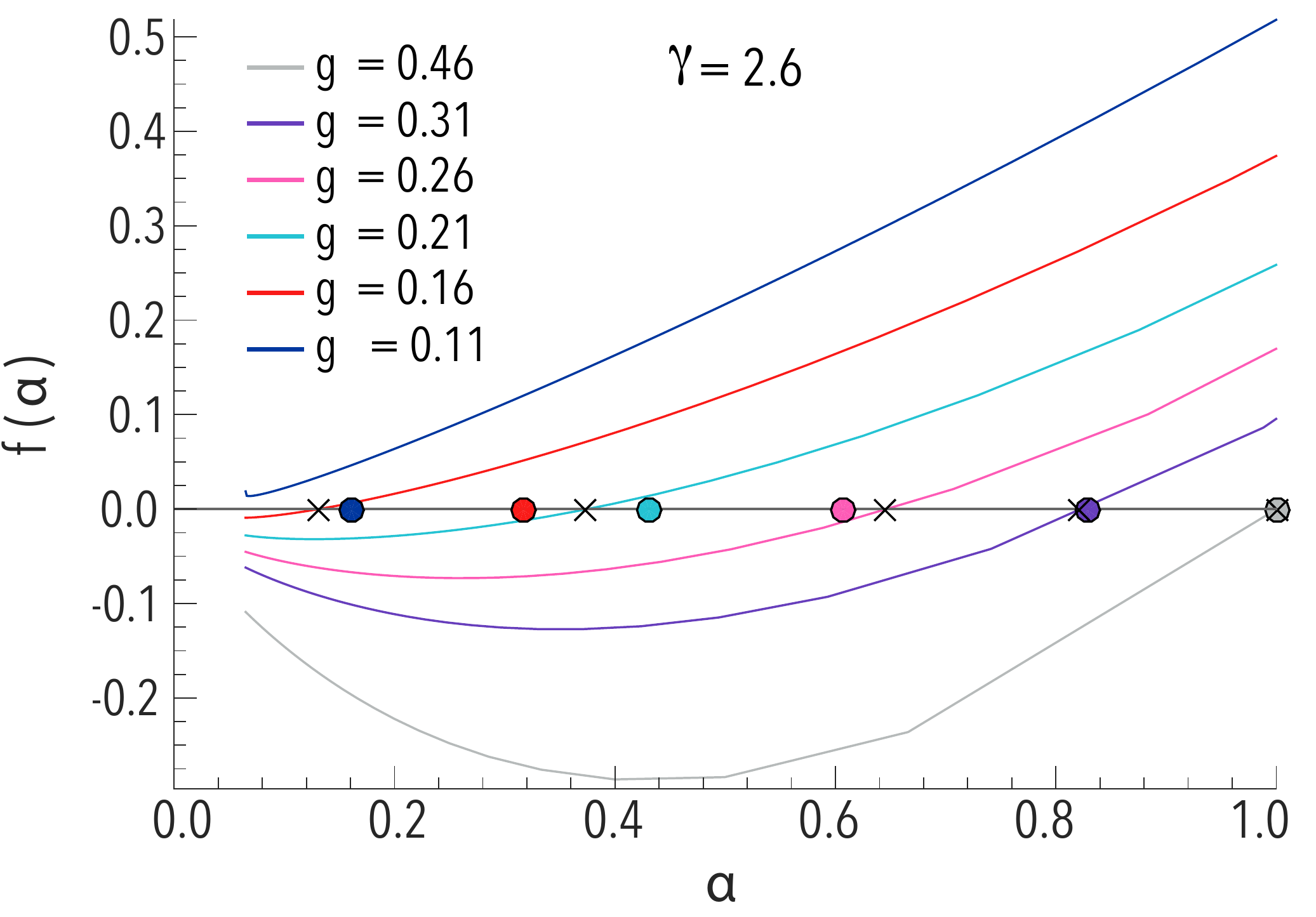}\\
		\includegraphics[height=50mm]{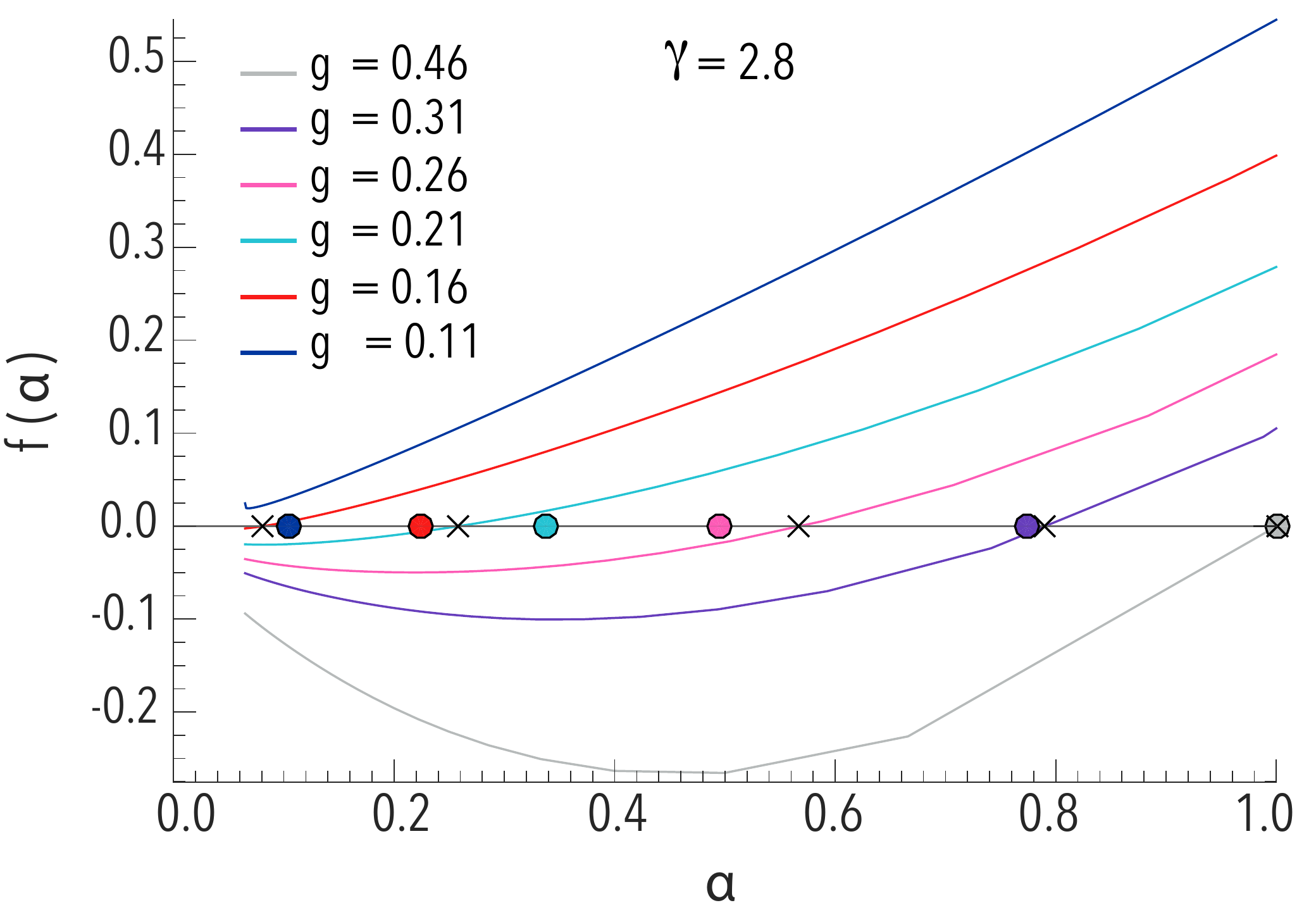}&
		\includegraphics[height=50mm]{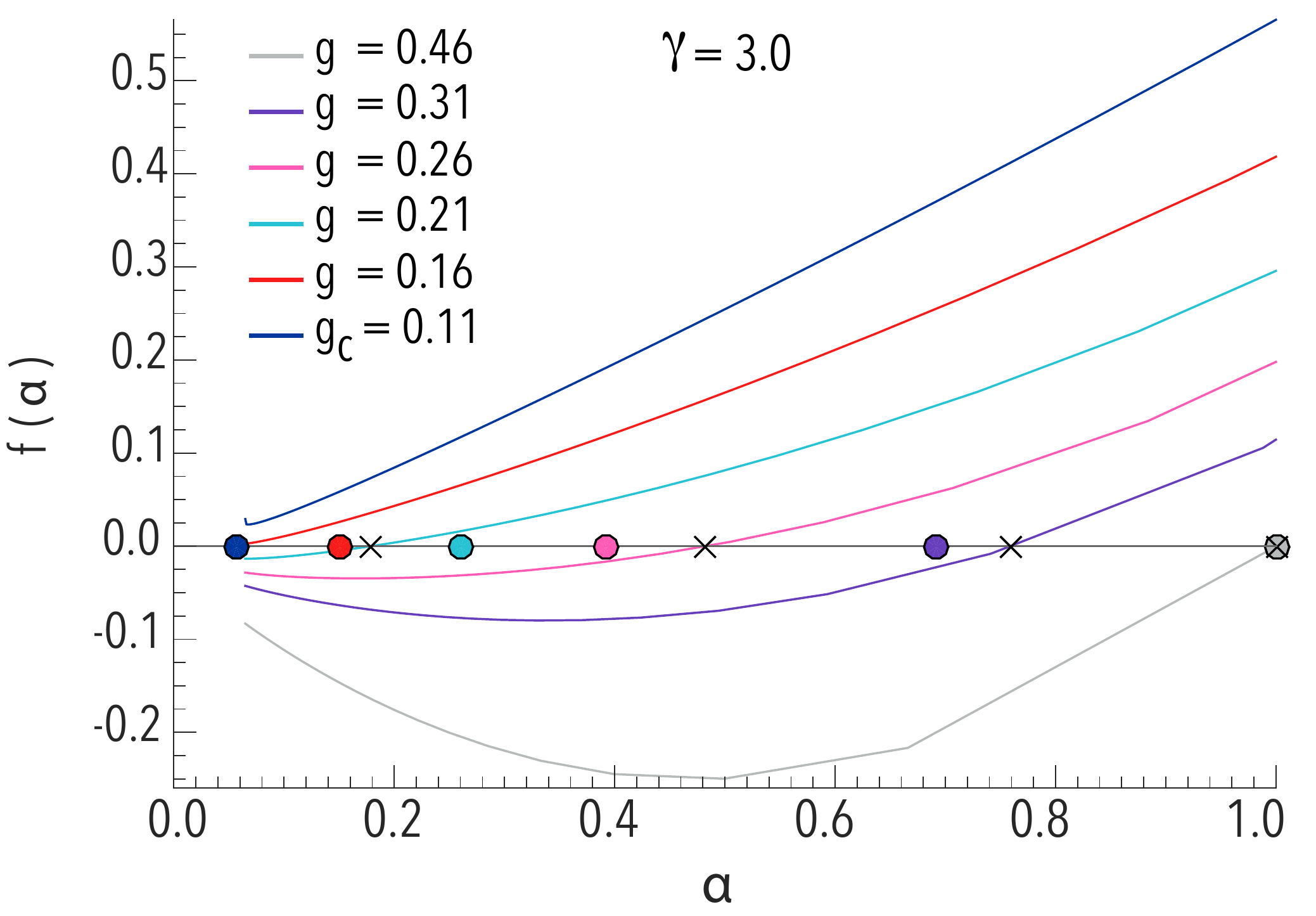}\\
	\end{tabular}
	\caption{\label{fig3.9} Theory performance. Each panel displays the function $f(\alpha)$ for several values of the coupling $g$ (constant $\forall i$). Black crosses indicate the roots of $f(\alpha)$ and, for comparison, solid circles represent the average firing rate computed from numerics (both, numerical results and analytical curves are coded by color). Note that for the higher values of $\gamma$ some analytical curves do not cross the x-axis (there is no -real- root). These cases are only for the smallest values of $g$. For larger values of $g$, it is possible to compute the roots and, thus, the average firing rate can be estimated. Parameters that were used are $N=5\times10^{4}$, $k_{min}=2$, $I_{ext}=0.85$, $\tau_{m}=10$, $\tau_{D}=1$ and $\theta=1.0$}
\end{figure*}
Note that $\alpha$ as a function of $g$ is constrained by \eqn~\ref{eqn3.12} and by the minimum degree of the network, so we know exactly where to look for --real-- roots of such equation. The value that \eqn~\ref{eqn3.12} yields for the upper bound of $g$ is $g_{sat}=\left[\theta-\left(1-e^{-\frac{\Delta t}{\tau_{m}}}\right)I_{ext}\right]/k_{min}$ (at which $\alpha=1$) and the lower bound of $g$ is given by $(\theta-I_{ext})/k_{min}$ (below this value neurons having $k=k_{min}$ cannot fire at all).\\
Now, the accuracy of our theory depends crucially on having enough statistics. For large values of $\gamma$ most terms in the RHS of \eqn~\ref{eqn3.17} do not contribute significantly (even for relatively low values of $k$), because they are quite small ($\sim N^{-1}$) compared to the other terms and, thus, only small connectivity groups dominate the predictions provided by \eqn~\ref{eqn3.17}. In accordance to the scale free model, the larger the value of $\gamma$ the less likely to have a significant amount of hubs to hold our approximations, so to have a sufficient number of nodes in most of $k$-classes that guarantee the contributions of most terms in the summation of \eqn~\ref{eqn3.17}, the value of $\gamma$ has to be reduced. \fig~\ref{fig3.9} shows evidence of this claim. In it, the behavior of $f(\alpha)$ is depicted for different couplings (six curves ranging from critical to saturation), and there is --at most-- one root for each of the values of $g$ considered.
It can be observed that for low values of $\gamma$ the roots of $f(\alpha)=0$ (represented by black crosses) are quite close to their corresponding average firing rates computed from simulations (solid circles), whereas for high values of $\gamma$ the theory is less accurate. In fact, our approach is quite precise when it comes to estimating the average firing rates for values of $g>\theta-I_{ext}$ when $\gamma \lesssim 2.5$ (e.g. for $\gamma=2.0$ the predicted values in this range of the pulse strength have $E_{\%}<5\%$ with respect to the average firing rates from numerics).\\
Regarding the roots of $f(\alpha)=0$ for $g \lesssim \theta-I_{ext}$ (above $g_c$, though not very far from it), no matter the value of $\gamma$, clearly our theory either underestimates such values or does not provide real roots to compare with the ones computed from simulations. The reason for this is related, again, to the topology of our system. In order to fire each neurons in the largest connectivity group ($k_{min}=2$) needs to receive excitatory inputs from all of their neighbors at the same time and when their membrane potentials is close to the resting value. Such conditions are fulfilled only occasionally by neurons that are not directly connected to hubs, which explains the large values of $\left\langle{\rm ISI}(k) \right\rangle_{t}$ already reported for small values of $k$ and $g$. This aspect of the model is responsible for increasing the variability of ISIs  of neurons having small degrees, and some of the assumptions in the derivation of \eqns~\ref{eqn3.10} and \ref{eqn3.17} may not be valid in this regime of pathological activity.\\

All in all, we believe that these findings are a great success because there are not many analytical results for models of neurons at the network level and, to our knowledge, none for leaky I\&F on sparse heterogeneous networks.

\section{Inhomogeneous couplings}

\begin{figure}[htp]
	\includegraphics[width=\columnwidth,clip=0]{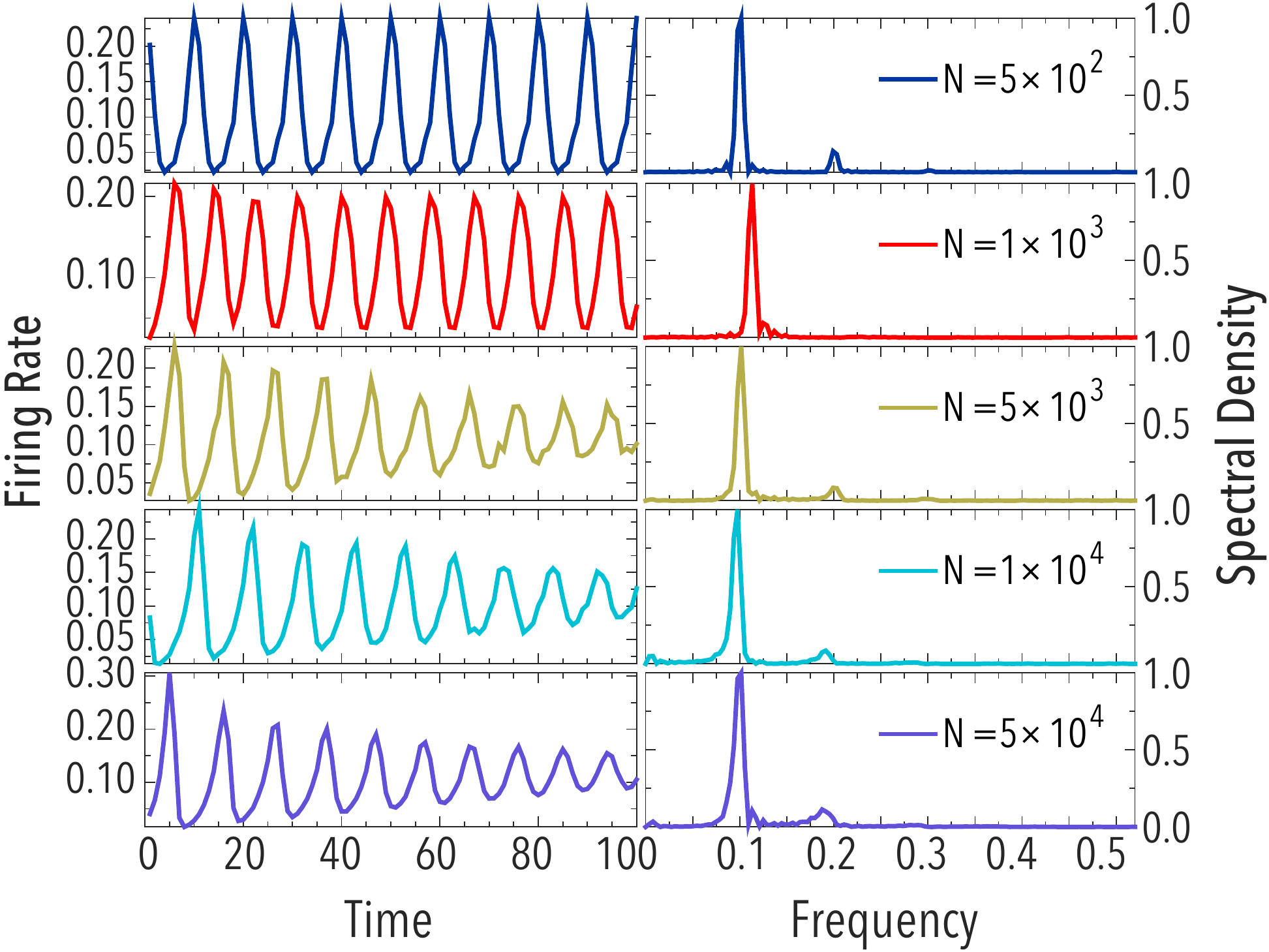}
	\caption{\label{fig3.10} Firing rate (left column) and the corresponding normalized spectral densities (right column) for five network sizes. Same parameters as before, except $g_{i}=0.90/k_{i}$.}
\end{figure}

\begin{figure}[htb]
	\includegraphics[scale=0.4,clip=0]{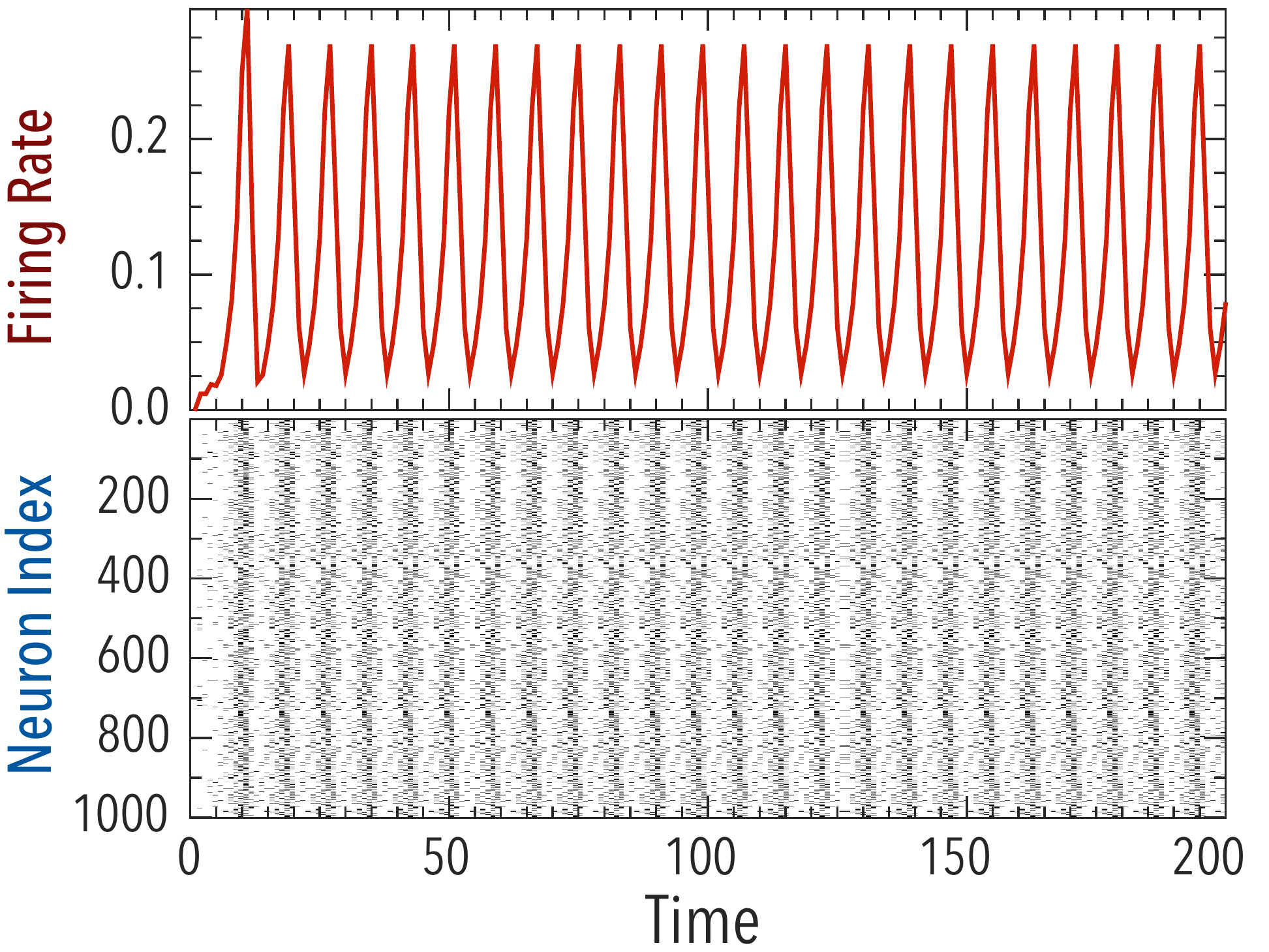}
	\caption{\label{fig3.11} Firing rate (top) and raster plot (bottom). Same parameters as before, except $g_{i}=0.89/k_{i}$.}
\end{figure}

\begin{figure}[htp]
	\includegraphics[scale=0.5,clip=0]{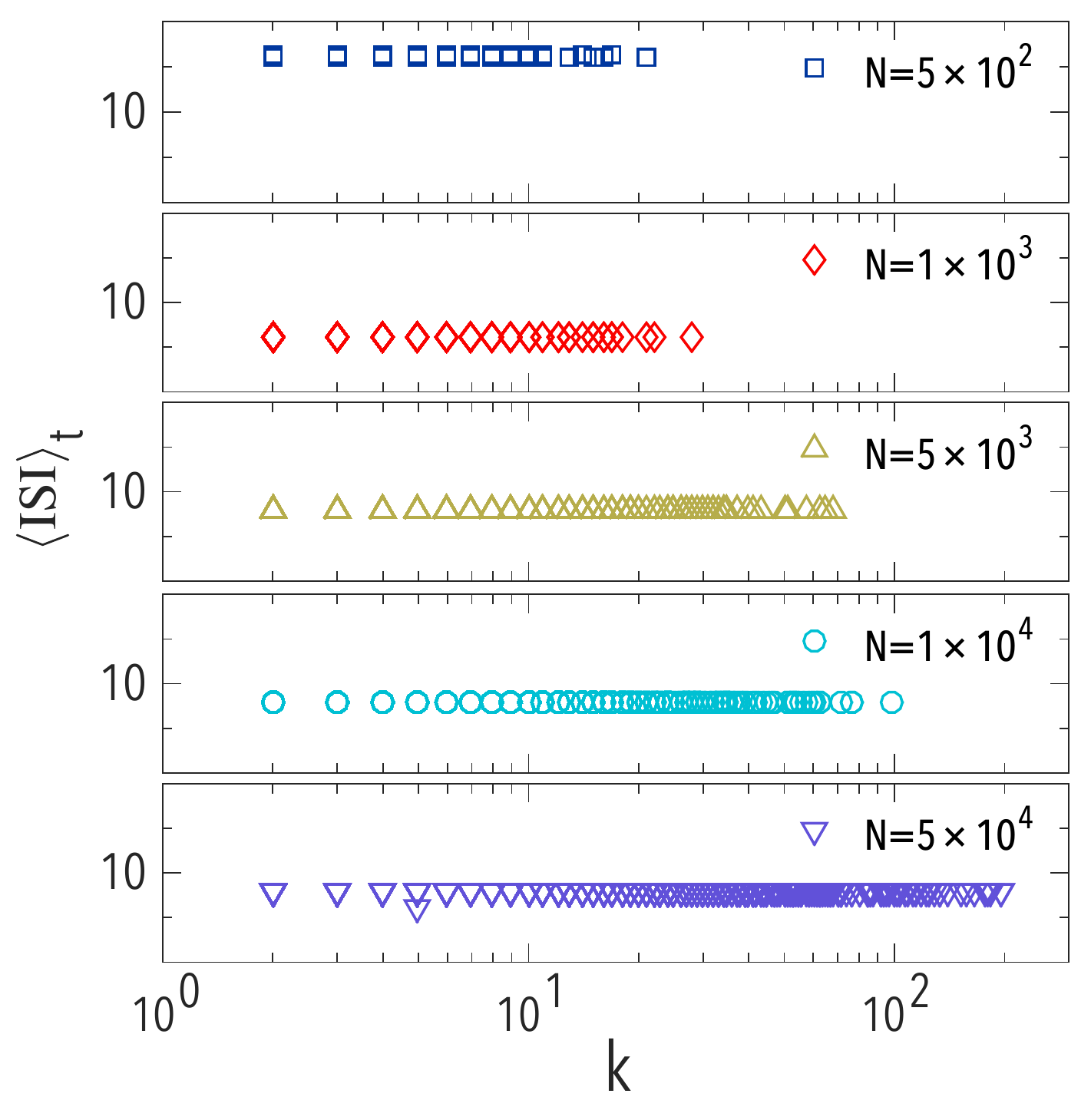}
	\caption{\label{fig3.12} Scatter plots of average ISI for five network sizes. Same parameters as before, except $g_{i}=0.89/k_{i}$.}
\end{figure}

\begin{figure}[htb]
	\includegraphics[scale=0.4,clip=0]{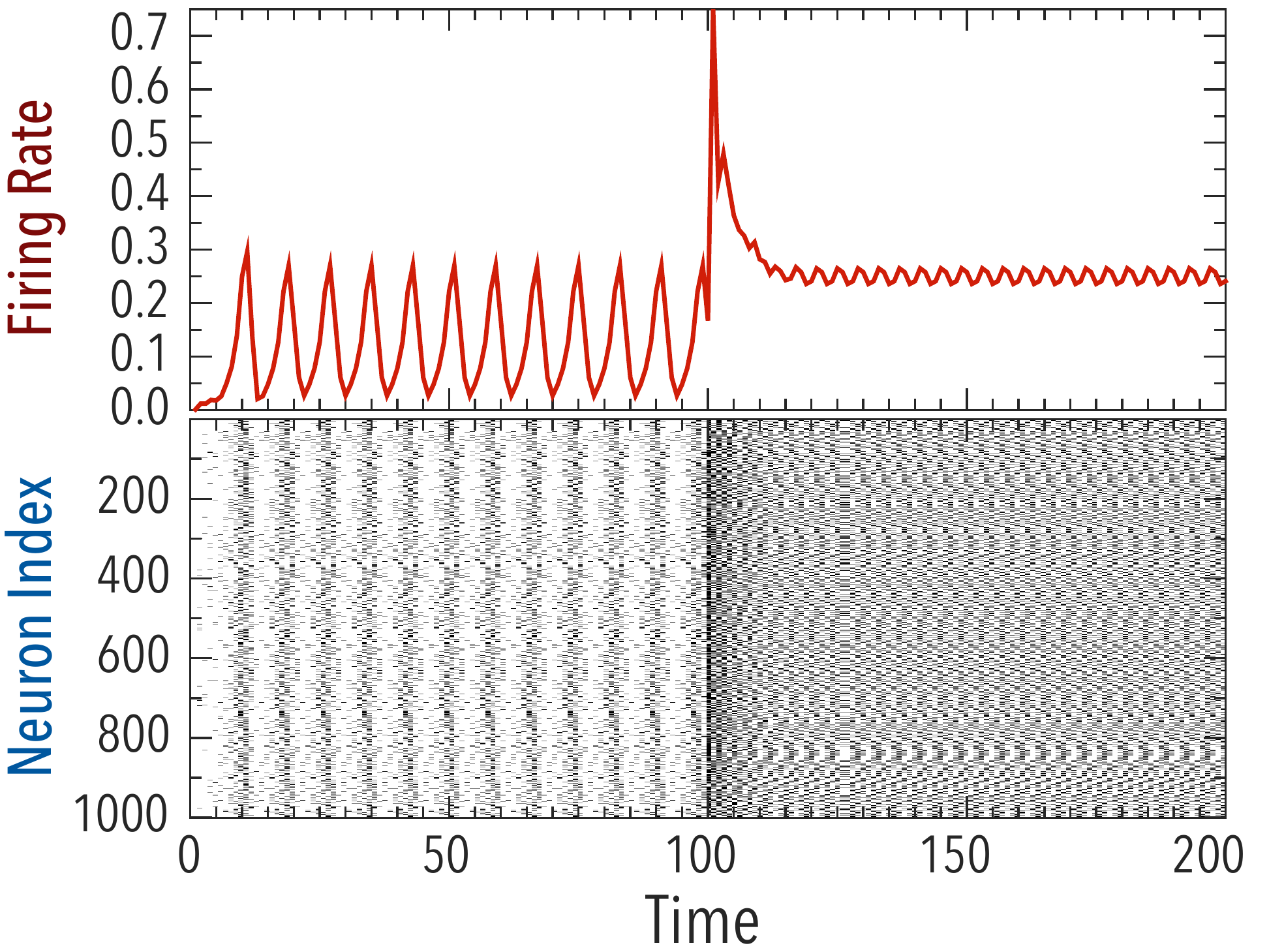}
	\caption{\label{fig3.13} Firing rate (top) and raster plot (bottom) for $g_{i}=0.9/k_{i}$. It is possible to change mean value, amplitude and structure (period and shape) by stimulating the ongoing dynamics with large excitatory signals.}
\end{figure}

If the strength of the pulses that each neuron receives is inversely proportional to its connectivity (i.e. $g_{i}~\propto~k^{-1}_{i}$, provided that the proportionality constant is above certain value --to obtain self-sustained activity), then the firing rate becomes periodic regardless of the size of the network and the initial firing conditions (as figure~\ref{fig3.10} shows). In this case the period achieved by the global signal is not necessarily small, as it usually is in saturation regime for homogeneous couplings, so the activity of the network seems to be an organized collective phenomenon of neurons. This result is essentially different from the periodic oscillations displayed in \fig~\ref{fig3.2}, not only because in this instance it is qualitatively the same for all the network sizes considered, but also because quantitatively both period and (proportional) amplitudes are similar for all cases as well. Note the amplitude of oscillations here $\sim 0.2N$, much greater than those exhibited by the same networks with homogeneous couplings ($\sim 0.05N$, at most). The difference in amplitudes is meaningful because it suggest that a significant amount of neurons (belonging to different $k$-classes) determine completely the structure of the global signal, and not just a small group with the same degree. Let us briefly comment a bit more this point. For inhomogeneous couplings most (or even all) neurons fire with similar average ISI (though not always the same instantaneous ISI), having only phase shifts between their signals. So the mechanism that explains self-sustained activity here is no longer a saturated cluster of neurons as it is for homogeneous couplings. As an average behavior, this can be confirmed by \eqn~\ref{eqn3.10} (because all the assumptions made in deriving it remain unchanged, granted that $g_i$ is a function exclusively of $k$). Thus $\left\langle{\rm ISI}(k) \right\rangle_{t}$ predicts the exact same average value for all $k$ precisely when $g_{i}~\propto~k^{-1}_{i}$. To illustrate this point, using the value of $\alpha$ from numerics (network of $N=10^{3}$), \eqn~\ref{eqn3.10} estimates that $\left\langle{\rm ISI}(k) \right\rangle_{t} \approx 7.1~({\rm arbitrary~time~units})\, \forall k$ and from spectral analysis we have determined that the period of the global firing signal is $8$ (arbitrary time units). \\
What these findings establish is that linking $g_{i}~\propto~k^{-1}_{i}$ is precisely one way to homogenize, dynamically, a system that is heterogeneous topologically (see reference \cite{Gomez2011} for another example of correlation between dynamics and topology). 
Unfortunately we cannot compute --$a~priori$-- a precise value of $\alpha$ from \eqn~\ref{eqn3.19}. Essentially, inhomogeneous couplings induce a form of initial firing conditions sensitivity, in the sense that the periodic signal reached by the network after the transient depends strongly on the specific procedure used to trigger the activity \cite{Cessac2007}. This effect is so sensitive that if we change the neurons at which the activity of a network starts, it generally leads to completely different firing rates, with differences in mean value, amplitude, periodicity and shape (see an example of this claim in \fig~\ref{fig3.12}). Obviously \eqn~\ref{eqn3.19} can not capture such diverse behaviors that occur for the same values of the parameters (however it may be useful to estimate some --not all-- of the average firing rates resulting from precise initial conditions).\\
\fig~\ref{fig3.11} illustrates another example of periodic oscillations, but in this instance all neurons fire exactly at the same frequency (statistically speaking). If the variance of $\{\left\langle{\rm ISI}(i) \right\rangle_{t}\}$, over the whole network, remains below $\sim10^{-4}$ for more than a certain length of the time window considered (which, in our case, is $10^{4}$ --arbitrary time units), we say that the system exhibits coherent oscillations (CO). For the outcome shown in \fig~\ref{fig3.11} this criterion is fully satisfied, so we conclude that the system self-organizes into one of many possible states of CO at the individual neuron level --though to achieve this state initial firing conditions have to be carefully selected. Note that CO can be interpret as a form of synchronization (in spite of the fact that these dynamical elements are not --intrinsically-- oscillators), thus we may call this behavior \emph{ISI synchronization}, which have also been detected in other systems of spiking neurons (limited to homogeneous topologies) \cite{McGraw2011,Olmi2012,Huerta2000,Karma1996}. Further evidence of the presence of CO for other network sizes is shown in \fig~\ref{fig3.12}, where the set of $N$ values of $\left\langle{\rm ISI}(i) \right\rangle_{t}$ are displayed. We observe that the system's variability is quite low, in fact, at relatively small values of $N$ (say, $N \lesssim 10^{3}$) the variance of $\{\left\langle{\rm ISI}(k) \right\rangle_{t}\}$ can be exactly zero (as predicted by \eqns~\ref{eqn3.10} and \ref{eqn3.11}) and for $\{\left\langle{\rm ISI}(i) \right\rangle_{t}\}$ it is around $10^{-4}$. Regarding larger networks ($N \gtrsim 10^{3}$), the variance of $\left\langle{\rm ISI}(i) \right\rangle_{t}$ is typically a bit higher (though still remains below $10^{-3}$), which can ve explained as a result of the so-called \emph{paradox of heterogeneity} (hubs are more difficult to synchronize).\\ From our extensive simulations, we observe that this regime of CO also exhibits longer transients that depend on the size of the system: \emph{the larger the system the longer the transient}. As an example, we have detected that transients can be as long as $10^{3}$ (arbitrary time units) for $N=5\times10^{4}$. However we suspect that in certain cases it might take longer to achieve full ISI synchronization (depending on the initial firing conditions and, of course, size of the system). For now, let us express that we have not determined yet all the ingredients that guarantee the appearance of CO in our model (just the coupling and careful selection of initial firing conditions), so ascertaining the rest remains a subject for future research. 
\\Finally, in order to test the response of the system for inhomogeneous couplings, we performed the same procedure introduced in \sect \ref{control}. As depicted in \fig~\ref{fig3.13} that after the excitatory stimulus is applied ($75\%$ of neurons at $t=100$) the mean value, the period and amplitude of the oscillations of the global signal are modified (but a periodic pattern remains). The activity of the network displays a major difference compared to previous cases seeing that we have here a full signal response. This example reveals that any stimulus applied to the SFNs under pulse heterogeneity might be globally irreversible, because there is no customary technique to restore the original state of the ongoing dynamics (but there might be a specific sequence to do it). Note also that the recovery time from the external stimulus is greater than in all previous results, which is consistent with the fact that hubs play a different roll for $g_{i}~\propto~k^{-1}_{i}$, as they are unable to spread the activity instantly like the do for homogeneous couplings. As for the effects of large inhibitory stimuli, we observe that global signal cease in all cases, as a result of reseting the membrane potential and the relative refractory period.
\\Summarizing our findings on inhomogeneous couplings:
\begin{itemize}
	\item Correlating $g_{i}~\propto~k^{-1}_{i}$ is good way to homogenize the activity of a network of excitable elements with heterogeneous topology.
	\item Even though the activity obtained is always well organized and can lead to the appearance of coherent oscillations, certainly it is very sensitive to initial firing conditions. 
	\item An external excitatory stimulus can alter mean value, period and shape of the global firing rate.
	\item As for the amplitude of the oscillations of the firing rate, large excitatory stimulus can reduce it significantly, but there is no general procedure to enhance the oscillations of the global signal by means of trivial external firing patterns.
	\item If the activity is initiated using extreme initial firing conditions (e.g. $N$ fires at the same time), it fails rapidly in all instances (the relative refractory period plays an important role here). Same reason explains why all inhibitory stimuli extinguish permanently the firing activity.
\end{itemize}

\section{Discussion}

The model studied in depth in this paper incorporates a leaky I\&F dynamics (with pulse-delays) on a scale-free structure of interactions. From the numerous computer simulations performed we note there are two regimes of self-sustain activity (for homogeneous coupling): First, when the pulse strength is greater than the voltage-gap between the firing threshold and the resting potential, one obtains what may be called \emph{standard} or \emph{normal} activity. In this state, all neurons in the system are able to fire whenever they receive one single input (regardless of the degree). The behavior of the global firing signal ranges from irregular patterns at relatively low values of the coupling to periodic ones at larger values (including the eventual full saturation of fires). Second, when the pulse amplitude is smaller than the voltage-gap between the firing threshold and the resting potential, then one obtains an \emph{abnormal} or \emph{pathological} activity, where most neurons in the smallest degree-class can only fire occasionally, when they receive inputs from $k_{min}$ sources at once. In this regime, the initial mechanism required to trigger the dynamics has, necessarily, to be extreme (i.e. involving multiple sources or even all nodes) and there is a value of the coupling, that we call \emph{critical}, defined as the smallest strength for which the population of excitable elements exhibits activity (lasting longer that certain time-window). This critical value is bound below by the voltage-gap mentioned above and the minimum degree of the network. The extreme firing initial conditions described above enables the system, at the critical coupling, to reach a global state of irregular fires where the mean firing rate is very small (which hints the possibility that the activity only involves certain parts of the network, those that are better connected). \\
Under this model, two distinct strategies to perturb the system were performed: First, {\bf dynamical perturbation:} An \emph{ad hoc} mechanism that involves the application of a large/global stimulus to the ongoing dynamics and observe its response. Particularly, we detected three scenarios; (i) inhibitory stimulus to cease all activity, (ii) no signal response to stimuli in the irregular-pattern regime, and (iii) the enhancement (hindrance) of oscillations' amplitude for a excitatory (inhibitory) stimulus in the regular-patterns regime (associated to strong couplings). Thus, we observed that there exists some external firing perturbations that may help to control the global activity by means of trivial mechanisms. Second, {\bf structural perturbation:} After performing a random removal of nodes (site percolation) the dynamics was executed in order to measure both the probability of activity failure and the average population firing rate. From this strategy one can confirm, once again, that scale-free networks are quite resilient. The threshold to activity failure depends on the size of the system as: \emph{the larger the network, the greater the fraction of removed nodes needed to achieve full activity failure}. Furthermore, we demonstrated that the average firing rate remained positive almost until the complete structural disintegration is achieved. \\
From some basic assumptions and with an elementary mathematical approach we managed to derive a few analytical expressions that allow us to predict diverse aspects of the dynamics. First of all, the average inter-spike-interval of the $k$-class proved to be an accurate method to decompose the dynamics and shed some light on the roles the different connectivity groups play. The activity ranges from saturation to critical behavior, and with this method we provided formal equations to compute the precise boundaries of the coupling strength and their corresponding average firing rates. A striking results of this formulation was the fact that at th critical value of the couping there is always a minimum saturated degree that can be computed analytically and it only depends on the single neuron parameters and the smallest degree of the network. Finally, we also derived an expression that links the mean firing rate with the average inter-spike-interval (as a function of $k$), which lead us to an implicit equation that can be solved without the recourse to simulations. This rigorous finding is accurate for standard activity on scale-free networks having high heterogeneity (i.e. lower values of the exponent of the degree distribution). The failure of the approximations on low heterogeneity graphs is a results of having poor statistics in most $k$-classes, whereas for abnormal activity one obtains unprecise estimations as a results of not having enough activity in the smallest-connectivity groups. \\
As mentioned at the beginning of this paper, most of the analytic results that can be found in the literature on models of neurons are focused either on single (isolated) units or on densely connected networks (to perform approximations impossible to achieve otherwise). Concerning the concept of sparseness of a network we should mention that is not always defined precisely or it is assumed vaguely by some authors of the field \cite{Brunel2000,Soula2006,Nicola2014}. For instance, in some of these studies the term ``sparse'' is used to describe recurrent networks (random regular graphs) in which each neuron receives $k=C \times N$ inputs (or connections). $C$ is simply a structural parameter (called connectivity level) to create an amorphous topology (random $k$--regular). However, $C\times N$ is typically assumed only one order of magnitude smaller than the size of the system ($C \sim 10^{-1}$) \cite{Amit1997a,Amit1997b,Mattia1998}. So, for example, when the structure has $N \sim 10^{5}$ nodes, then each of them is connected to $k \sim 10^{4}$ neighbors (and additional connections from outside the network are also allowed). Clearly this is not what we commonly know as a sparse graph in network science \cite{Genio2011}. In spite of all the limitations found in our theory, the fact that our analytics works on sparse heterogeneous networks (where, for instance, $\left\langle k \right\rangle \sim 10^{1}$, quite small compared to $N\sim 10^{5}$) is a significant accomplishment.\\
Another prominent aspect of \eqns~\ref{eqn3.10}, \ref{eqn3.11} and \ref{eqn3.17}, that we want to strongly emphasize, is that these equations are not valid exclusively for SFNs but for any topology with heterogeneous degree distribution (provided that all the other premises are satisfied as well). Using the scale-free model in the present paper can be interpret as an extreme example of such heterogeneity, but there is no constraint that attaches our finding to this type of network at all, and applying the same framework to other kinds of topologies (for the same dynamics) should produce accurate results as well. \\
As for our last result, we analyzed the activity under inhomogeneous couplings. As predicted by our analytics, correlating the pulse-strength of the inputs that a neuron receives with its degree is a way to homogenize the activity of a network of excitable elements with heterogeneous topology: The activity become well-organized and periodic, all neurons exhibit similar inter-spike-interval. However, under this assumption, the system becomes sensitive to initial firing conditions. Different mechanisms to trigger the dynamics lead to distinct activity patterns. As an extreme instance of this behavior, one can obtain a type of synchronization (named coherent oscillations) by carefully selecting initial firing conditions, though further research is needed on this phenomenon.\\
We believe that the study of spiking models is not restricted to traditional Neuroscience, and that the measure of success of such models should not be limited as to how accurate they are to reproduce what is observed in live neurons. There are many real systems where agents do not interact continuously, but in an intermittent fashion, and the lack of applications of these neural models constitutes a great opportunity to formulate novel research in the view of new progress.

\bibliography{Refs}        
\bibliographystyle{unsrt}  

\end{document}